\documentclass[printer]{aa}

\usepackage{graphicx}
\usepackage{txfonts}

\usepackage{amssymb}
\usepackage{mathtools}
\usepackage{lscape}
\usepackage{natbib}
\bibpunct{(}{)}{;}{a}{}{,} 
\usepackage{gensymb}
\usepackage{amsmath}
\usepackage{adjustbox}
\usepackage{multirow}
\usepackage{epstopdf}
\usepackage{tabulary}
\usepackage{tabularx}
\usepackage{hhline}
\usepackage{wasysym}
\usepackage{caption}
\begin{document}

   \title{Low-velocity collision behaviour of clusters composed of sub-mm sized dust aggregates}

   \author{J. Brisset
          \inst{1} \inst{2} \inst{3}
          \and
          D. Hei\ss elmann \inst{1}
	\and
	S. Kothe \inst{1}
	\and
	R. Weidling \inst{1}
	\and
	J. Blum \inst{1}
          }

   \institute{Institut fuer Geophysik und extraterrestrische Physik, Technische Universitaet Braunschweig, Mendelssohnstr. 3, 38106 Braunschweig, Germany\\
              \email{j.brisset@tu-bs.de}
         \and
             Max Planck Institute for Solar System Research, Justus-von-Liebig-Weg 3, 37077 Goettingen, Germany
         \and
	    Present address: centre of Microgravity Research, University of Central Florida, 4111 Libra Drive, Orlando FL-32816, USA
             }


 
  \abstract
   {The experiment results presented apply to the very first stages of planet formation, when small dust aggregates collide in the protoplanetary disc and grow into bigger clusters. In 2011, before flying on the REXUS 12 suborbital rocket in 2012, the Suborbital Particle and Aggregation Experiment (SPACE) performed drop tower flights. We present the results of this first microgravity campaign.}
   {The experiments presented aim to measure the outcome of collisions between sub-mm sized protoplanetary dust aggregate analogues. We also observed the clusters formed from these aggregates and their collision behaviour.}
   {The experiments were performed at the drop tower in Bremen. The protoplanetary dust analogue materials were micrometre-sized monodisperse and polydisperse SiO$_2$ particles prepared into aggregates with sizes between 120~$\mu$m and 250~$\mu$m. One of the dust samples contained aggregates that were previously compacted through repeated bouncing. During three flights of 9~s of microgravity each, individual collisions between aggregates and the formation of clusters of up to a few millimetres in size were observed. In addition, the collisions of clusters with the experiment cell walls leading to compaction or fragmentation were recorded.}
   {We observed collisions amongst dust aggregates and collisions between dust clusters and the cell aluminium walls at speeds ranging from about 0.1~cm~s$^{-1}$ to 20~cm~s$^{-1}$. The velocities at which sticking occurred ranged from 0.18 to 5.0~cm~s$^{-1}$ for aggregates composed of monodisperse dust, with an average value of 2.1$\pm$0.9~cm~s$^{-1}$ for reduced masses ranging from 1.2$\times10^{-6}$ to 1.8$\times10^{-3}$~g with an average value of 2.2$^{+16}_{-2.1}\times10^{-4}$~g. The velocities at which bouncing occurred ranged from 1.9 to 11.9~cm~s$^{-1}$ for the same aggregates with an average of 5.9$\pm$3.2~cm~s$^{-1}$ for reduced masses ranging from 2.1$\times10^{-6}$ to 2.4$\times10^{-4}$ with an average of 7.8$\pm$2.4$\times10^{-5}$~g. The velocities at which fragmentation occurred ranged from 4.9 to 23.8~cm~s$^{-1}$ for the same aggregates with an average of 10.1$\pm$3.2~cm~s$^{-1}$ for reduced masses ranging from 1.2$\times10^{-5}$ to 1.2$\times10^{-3}$ with an average value of 4.2$\pm$2.4$\times10^{-4}$~g. From the restructuring and fragmentation of clusters composed of dust aggregates colliding with the aluminium cell walls, we derived a collision recipe for dust aggregates ($\sim$100~$\mu$m) following the model of Dominik \& Thielens (1997) developed for microscopic particles. We measured a critical rolling energy of 1.8$\pm0.9\times10^{-13}$~J and a critical breaking energy of 3.5$\pm1.5\times10^{-13}$~J for 100~$\mu$m-sized non-compacted aggregates.}
   {}


\maketitle

\section{Introduction}
\label{s:intro}

The latest progress in observational techniques opens promising prospects on a better understanding of planet formation. The Kepler space telescope is continuously discovering new planets orbiting other stars \citep{howard_et_al2012ApJS}, indicating that the processes leading to the formation of planets are actually very common in our universe. Furthermore, the revolutionary resolution in sub- and mm-wavelengths obtained with the ALMA telescope (Atacama Large Millimetre Array in Chile) now allows for a direct observation of the presence and spatial distribution of mm- to cm-sized dust grains in protoplanetary discs (PPDs) down to a few astronomical units \citep[e.g.][]{macgregor_et_al2013APJL,tazzari2016aap}. As these discs harbour planet formation, these observations are now delivering new insights into the processes leading to planetary systems similar to our own. Owing to the very long timescales of astronomical processes \citep[discs evolving on a timescale of $\sim10^6$~years;][]{hernandez_et_al2007ApJ,yasui_et_al2012AAS,ribas2015aap}, observations alone are not sufficient to get a complete picture of the processes leading to planet formation. The combination of observations with theoretical models, numerical simulations and experimental studies is essential for an overall understanding. \\
In a PPD, two particles that are colliding can stick together because of short range dipole-dipole interactions between molecules, known as the Van der Waals force. This attractive force can be modelled as a specific surface energy at the contact area between the particles holding them together \citep{johnson_et_al1971PRSL}. It has to be overcome to separate two particles sticking to each other and the outcome of a collision depends on the collision energy. This is why low-velocity collisions between small particles always lead to sticking \citep{blum2000_PRL,poppe_et_al2000bApJ}. Bigger and more complex aggregates (composed of a large number of monomer particles), however, are influenced by interactions with the surrounding gas and obtain higher relative velocities, leading them to leave the "hit and stick" regime \citep[e.g. see the reviews and collision model by][]{johansen_et_al2014PP,testi_et_al2014PP,blum_wurm2008ARAA, guettler_et_al2010A&A}. Therefore, direct agglomeration of $\mu$m-sized dust particles into km-sized planetesimals via simple growth by sticking appears challenging.\\
In addition to experiments on $\mu$m-sized particles \citep[e.g.][]{poppe_et_al2000aApJ,poppe_et_al2000bApJ,gundlach_blum2015ApJ} and small aggregates thereof \citep{blum2000_PRL, blum_wurm2000_icarus}, several dust collision experiments were performed by observing aggregates of higher masses and at higher relative velocities (see the review by \cite{blum_wurm2008ARAA}, and the collision model by \cite{guettler_et_al2010A&A}). Several of these experiments were combined with the Hertz and \cite{thornton_ning2001} contact theories \citep[e.g. by ][]{guettler_et_al2010A&A,kothe_et_al2013Icarus} resulting in a dust collision model. This model predicts the outcome of a collision between two dust particles as a function of their mass and relative velocity; for certain aggregate masses, increasing relative velocities lead to a transition in collision outcomes from sticking to bouncing and fragmentation for even higher velocities. These transitions with increasing particle mass and collision velocity are also observed in molecular dynamics simulations \citep[see e.g.][]{dominik_and_tielens1997ApJ,wada_et_al2009ApJ}.\\
The impact of the sticking to bouncing transition becomes apparent when it is taken into account in PPD dust growth models and numerical simulations. In Monte Carlo dust collision simulations of \citet{zsom_et_al2010AA} for example, the introduction of aggregate bouncing stalls their growth at sizes around 1~cm in diameter (bouncing barrier). In \cite{brauer_et_al2008AAa}, aggregate fragmentation that stalls the growth and keeps the dust size distribution under a few mm in radius (fragmentation barrier). These barriers imply that dust agglomeration to planetesimal sizes requires more complex processes than simple growth upon collision. Information about aggregate collision behaviour at masses and velocities around the transitions between sticking and bouncing and between bouncing and fragmentation are therefore very important for the further investigation of dust growth in PPDs. The present work concentrates on the sticking to bouncing transition for 100~$\mu$m-sized aggregates. One possible scenario in which these aggregates may become building blocks for the growth of larger clusters in the PPD is their formation in a high-velocity environment, such as turbulent zones, and the subsequent drift to quieter areas of the disc in which their relative velocities are smaller. The reduced collision speeds after such a drift would lead to the resumption of aggregate growth from 100~$\mu$m-sized constituents.\\
The Suborbital Particle Aggregation and Collision Experiment (SPACE) is part of a series of experiments attempting to better define the parameters leading to aggregate sticking, bouncing, or fragmenting upon collision at velocities of 10~cm~s$^{-1}$ and below. One of the challenges of observing dust collisions in many particle systems at such gentle velocities is the necessity to conduct these experiments under microgravity conditions. In addition to flying on the REXUS~12 suborbital rocket ($\sim$150~s of consecutive microgravity time) in 2012 \citep{brisset_et_al2016AA}, the SPACE experiment also performed a flight campaign at the Bremen drop tower in 2011. This work presents the experimental results obtained from these $\sim$9~s microgravity experiments. Section~\ref{s:setup} describes the hardware set-up, experiment run parameters, and dust samples investigated. In Section~\ref{s:results}, the results obtained from the experiment runs are presented. Section~\ref{s:model_input} shows how the data results provide input for the current dust collision model. Section~\ref{s:discussion} draws conclusions on sub-mm-sized dust aggregate collision behaviour from the experimental data obtained.

\section{Experimental set-up}
\label{s:setup}

This section presents the experiment set-up that was used at the Bremen drop tower in August 2011. More details about the hardware can be found in \cite{brisset_et_al2013RSI}.

\paragraph{\textit{Hardware}}

The Bremen drop tower offers 4.74~s of microgravity time when dropping an experiment capsule and up to 9.3~s when using the catapult. Disturbances to the microgravity quality are lower than 10$^{-6}g$, where $g$ is the gravitational acceleration of the Earth. The camera on board the capsule is a Photron Fastcam MC2, which produces 8-bit greyscale frames and records at a rate of 500~fps and a resolution of 512 $\times$ 512~px (optical resolution of 57 $\mu$m/px).\\
The experiment set-up consists of four glass experiment cells with identical dimensions of 10.2 $\times$ 7.6 $\times$ 12.7~mm$^3$ (see Figure~\ref{f:dt_cells}). The glass used was common soda lime glass cut out of microscope slides, which had been anti-adhesively coated by the Frauenhofer Institute for Surface Engineering and Thin Films of Braunschweig \citep[see][ and their Figure~5]{brisset_et_al2013RSI}.The four experiment cells were filled with several samples of SiO$_2$ dust (see the following paragraph for details) and agitated in a circular manner to trigger aggregate collisions and agglomeration under microgravity conditions. In total, the experiment flew five times during the drop tower campaign.\\
\vspace*{-\baselineskip}

\begin{figure}[t]
  \begin{center}  
  \includegraphics[width=0.45\textwidth]{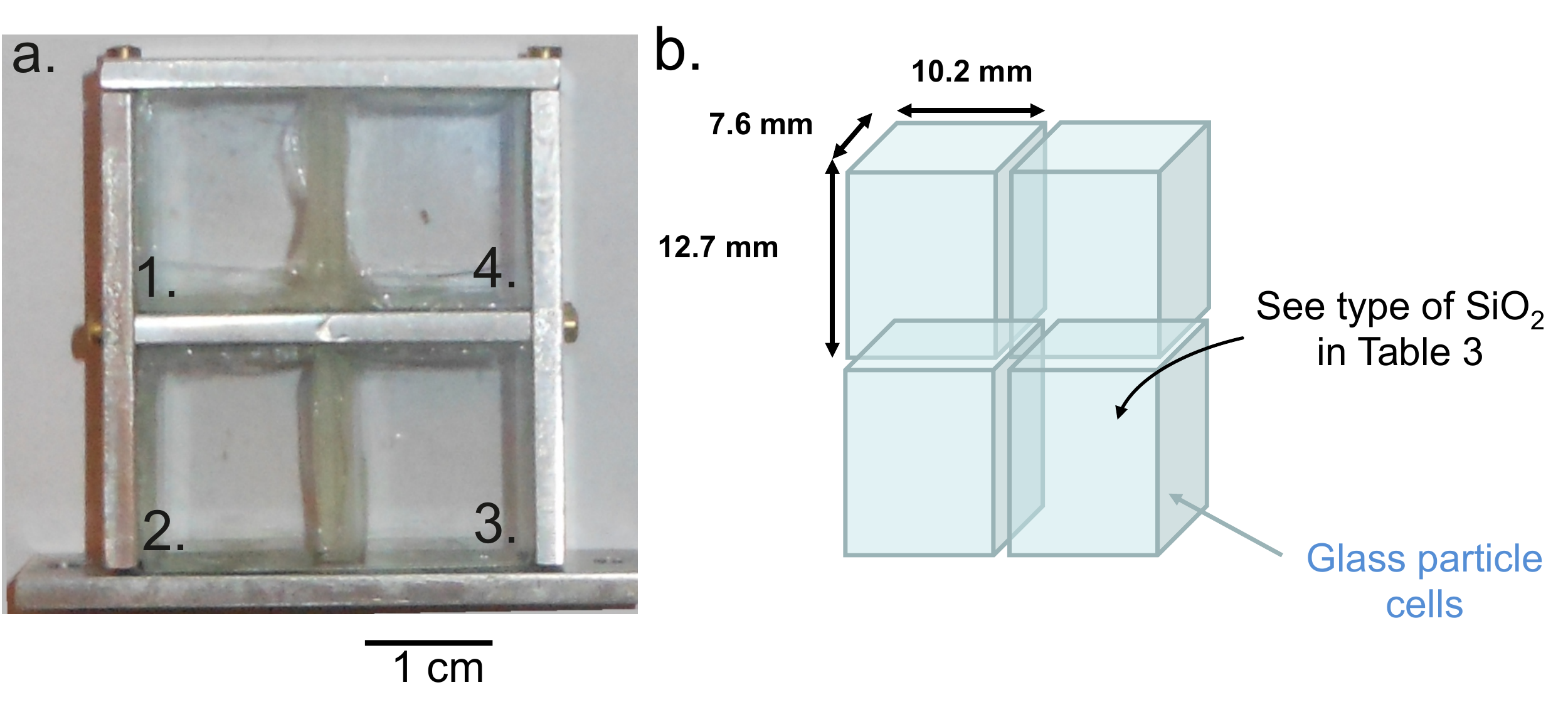}\\
 \caption{Representation of experiment sample cells. a) Picture of the glass cells numbered from 1 to 4. b) Schematic showing the four glass cells with their internal dimensions indicated.}
 \label{f:dt_cells}
 \label{f:dt_cells}
 \end{center}
\end{figure}

\paragraph{\textit{Dust samples}}

\begin{table*}[t]
\begin{center}
  \caption{Details of the SiO$_2$ aggregate types and sizes observed. The asterisk (*) denotes spherical polydisperse dust manufactured by Admatechs (type SO-E3). The other polydisperse dust listed is irregular and manufactured by Sigma-Aldrich (Table~\ref{t:silicates}). As these experiments were performed with a test version of the experiment hardware \citep[later to be flown on a suborbital rocket; see][]{brisset_et_al2016AA}, not all cells were filled with dust aggregates and not all drops delivered useful experimental data. The agglomerate sizes are indicated with their standard deviation. The compaction of the monodisperse aggregates during drop~3 is described in the text.}
    \begin{tabular}{|>{\centering}p{0.37in}|>{\centering}p{0.3in}|>{\centering}p{0.92in}|>{\centering}p{0.3in}|>{\centering}p{1in}|>{\centering}p{0.75in}||>{\raggedleft\arraybackslash}p{1.45in}|}
    \hline
    \textbf{Drop} & \textbf{Cell} & \textbf{Dust type} & \textbf{Code} & \textbf{Mean aggregate diameter [$\mu$m]} & \textbf{Quantity [mg]} & \textbf{Result type} \\
    \hline
    \multirow{2}[8]{*}{2} & 1     & polydisperse & SP & 156$\pm$28  & 11.0  & no agglomeration \\
\cline{2-6}          & 2     & polydisperse* & IP & 156$\pm$28  & 11.0  & monomer agglomeration \\
\cline{2-6}          & 3     & monodisperse & M & 118$\pm$88 & 14.3  & cluster agglomeration \\
    \hline
    \multirow{2}[8]{*}{3} & 2     & polydisperse & SP & 156$\pm$28  & 11.7  & no agglomeration \\
\cline{2-6}          & \multirow{2}[2]{*}{3} & compacted  & CM & 118$\pm$88  & 6.8   & \multirow{2}[2]{*}{cluster agglomeration} \\
          &       & monodisperse &    &   &       &  \\
    \hline
    \multirow{2}[7]{*}{5} & 1     & polydisperse* & IP & 156$\pm$28  & 8.8   & cluster agglomeration \\
\cline{2-6}          & 2     & monodisperse & M & 118$\pm$88  & 6.8   & cluster agglomeration \\
\cline{2-6}          & 3     & monodisperse & M & 118$\pm$88  & 6.8   & cluster agglomeration \\
    \hline
    \end{tabular}
  \label{t:dt_dust}
\end{center}
\end{table*}

\begin{center}
\begin{figure*}[hbtp]
    \begin{minipage}{0.5\textwidth}
      \begin{center}
       \includegraphics[width=0.75\textwidth]{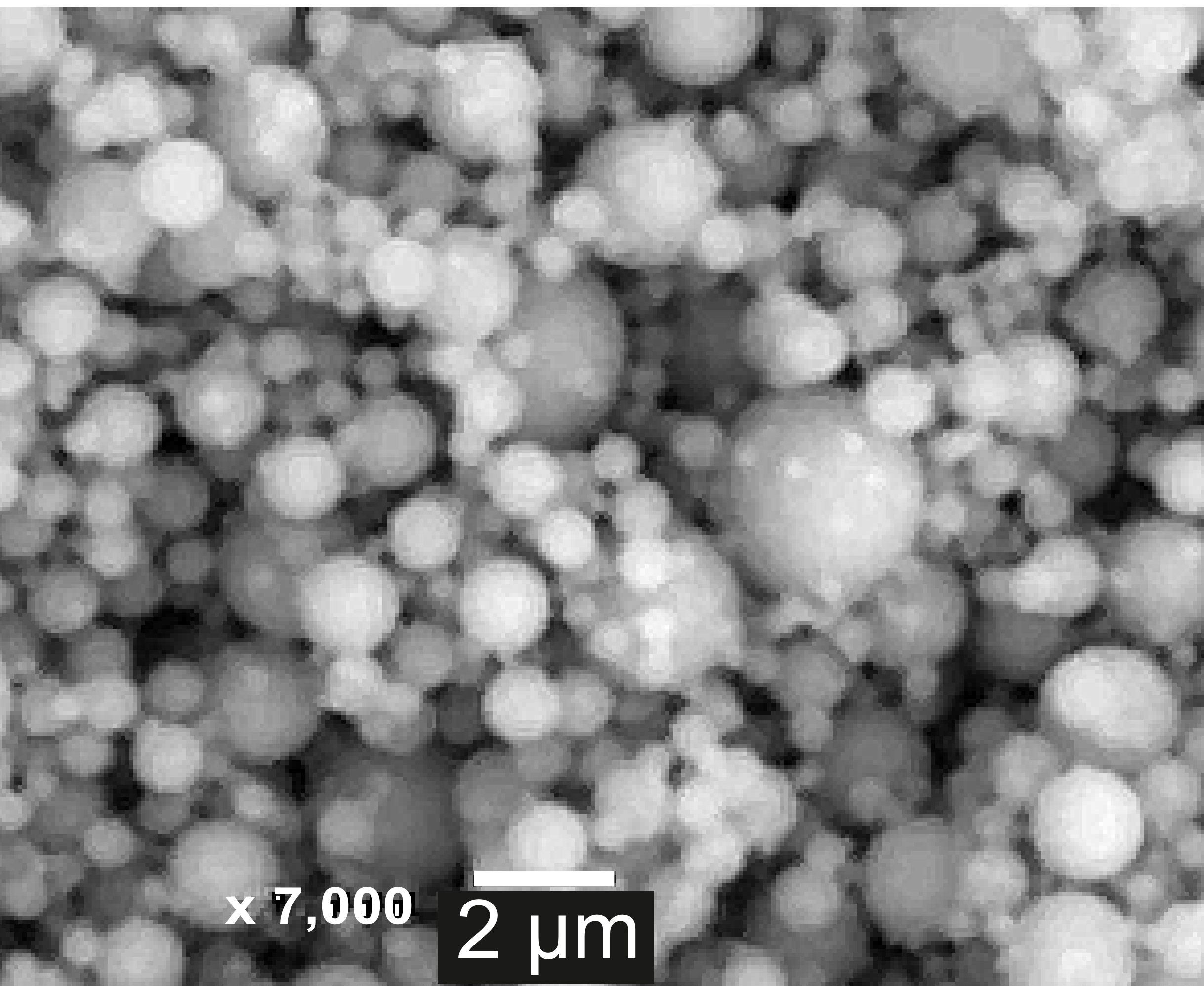}
        \setcaptionwidth{0.7\textwidth}
 	\caption{SEM picture of an aggregate composed of polydisperse SiO$_2$ (Admatechs). Image credit: N. Machii.}
	 \label{f:poly_pic}
      \end{center}
    \end{minipage}
    \begin{minipage}{0.5\textwidth}
      \begin{center}
      	\includegraphics[width=1\textwidth]{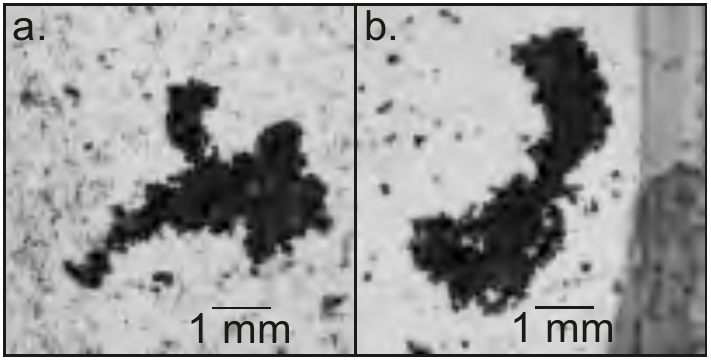}
      	\setcaptionwidth{1\textwidth}
	 \caption{Examples of clusters formed during experiment runs. a) Cluster formed in cell 1 after 7.9 s of drop 5 (aggregates composed of spherical polydisperse SiO$_2$, see Table \ref{t:dt_dust}) is shown. b) Cluster formed in cell 2 after 8.5 s of drop 5 is shown (aggregates composed of monodisperse SiO$_2$, see Table \ref{t:dt_dust}). A movie of this particular experiment is available  online.}
      \label{f:clusters}
      \end{center}
    \end{minipage}
\end{figure*}
\end{center}

In this paper, we refer to the single dust aggregates (a few 100 $\mu$m in size) introduced into the experiment glass cells before the beginning of the experiment as aggregates, dust aggregates, or monomer aggregates. These aggregates themselves consist of smaller dust grains of $\sim$1 $\mu$m in size, which we call monomer particles. In their storage container, these monomer particles form aggregates, which we then sieved to the desired size distributions. When several of these dust aggregates stick together during the experiment to form a bigger agglomerate, we refer to this as a cluster.\\
The SiO$_2$ aggregates prepared were sieved into a size distribution between 100~and~250~$\mu$m and were distributed amongst the experiment cells as listed in Table \ref{t:dt_dust}. Four different types of aggregates were used. The first consisted of monodisperse dust, composed of monodisperse spherical particles of 0.76~$\mu$m radius, manufactured by Micromod \citep[see Figure 3b. in][]{brisset_et_al2016AA}. The second were compacted aggregates of monodisperse dust. These aggregates were prepared as those  mentioned above but in addition, they were shaken on a metal plate for 10 min at 10 Hz (plate shaking frequency). This leads to a compaction of the outer parts of the aggregates \citep{weidling_et_al2009ApJ}. The third type were aggregates of polydisperse dust, which is composed of irregularly shaped particles with radii between 0.05~$\mu$m and 5~$\mu$m, manufactured by Sigma-Aldrich \citep[see Figure 3a. in][]{brisset_et_al2016AA}. Finally, the last aggregate type were aggregates of polydisperse dust consisting of polydisperse spheres with radii between 0.8~$\mu$m and 1.2~$\mu$m, manufactured by Admatechs (type SO-E3, see Figure \ref{f:poly_pic}). The properties of these dust samples are listed in Table~\ref{t:silicates}. The volume filling factor of the sieved clusters was measured to be $\phi$=0.37$^{+0.06}_{-0.05}$ \citep{weidling_et_al2012Icarus,kothe_et_al2013Icarus}. For the previously compacted aggregates, we used the method described in \cite{weidling_et_al2009ApJ} and our aggregates mass and volume values (10$^{-5}$~g and 4.2$ \times10^{-12}$~m$^3$, respectively) to calculate the volume filling factor of the compacted aggregate rim. We found a value of 0.54 instead of a uniform 0.37 for the non-compacted aggregates.\\
The choice of the aggregate size distribution around $\sim100~\mu$m was motivated by the region of the parameter space to be studied in the dust collision model. The transition from sticking to bouncing has already been well studied for smaller and larger aggregates \citep{blum_wurm2008ARAA,guettler_et_al2010A&A}. As $\sim100\mu$m-sized aggregates can also serve as building blocks for the growth of clusters (see Section~\ref{s:size_relevance}), the experiment presented here concentrated on these aggregate sizes in particular.

\begin{table*}[t]
  \begin{center}
  \caption{Properties of the dust types investigated.}
    \begin{tabular}{|l|c|c|c|}
    \hline
    \textbf{Dust type} & Spherical & Spherical & Irregular \\
    			& Monodisperse & polydisperse & polydisperse \\ \hline
    \textbf{Manufacturer} & Micromod & Admatechs (SO-E3) & Sigma-Aldrich \\ \hline
    \textbf{Monomer radius ($\mu$m)} & 0.76 $\pm$ 0.03$^{(1)}$ & 0.8 to 1.2$^{(3)}$  & 0.05 - 5$^{(5)}$ \\ \hline
    \textbf{Density (kg/m$^3$)} & 2000$^{(1)}$  & 2600$^{(3)}$  & 2600$^{(5)}$ \\ \hline
    \textbf{Young's modulus (GPa)} & 54$^{(2)}$  & 41$^{(4)}$  & 41$^{(4)}$ \\ \hline
    \textbf{Poisson number} & 0.17$^{(2)}$  & 0.17$^{(4)}$  & 0.17$^{(4)}$ \\ \hline
    \end{tabular}
    \label{t:silicates}
    \end{center}
   $^{(1)}$ manufacturer information Micromod\\
   $^{(2)}$ \cite{seizinger_et_al2012AA}\\
   $^{(3)}$ manufacturer information Admatechs\\
   $^{(4)}$ \cite{weidling_et_al2012Icarus}\\
   $^{(5)}$ manufacturer information Sigma-Aldrich GmbH
\end{table*}

\paragraph{\textit{Shaking profile}}

The shaking of the experiment cells is necessary to allow for the continuous generation of collisions between the aggregates during a drop. Without continuous agitation, collisions between aggregates tend to decrease their relative velocities on a very short timescale \citep{heisselmann_et_al2007}.\\
The amplitude of the rotational motion used for shaking was 1~mm. The shaking frequencies of each experiment are listed in Table \ref{t:dt_sh_profile}. The first few seconds were used for de-agglomerating clusters that formed during the experiment preparation and the capsule launch. The shaking frequencies of these first seconds ranged from 10.5~Hz to 16.7~Hz. This fast shaking phase was followed by a slower shaking phase, with shaking frequencies from 3~Hz to 6~Hz, allowing for low-velocity collisions and cluster formation (see Section~\ref{s:dt_collisions}).\\

\paragraph{\textit{Resulting data}} During drops 2, 3 and~5, the formation of clusters of sizes up to 5~mm could be observed in some of the cells. The irregular polydisperse dust did not agglomerate into bigger clusters (drops 2 and~3). The spherical polydisperse dust agglomerated only a few monomer aggregates at a time at high shaking frequencies (drop~2) and formed larger clusters (up to $\sim$1000~monomer aggregates) at lower shaking frequencies (drop~5). The monodisperse dust, whether compacted or not, formed large clusters during all three drops. These clusters formed free-flying inside of the experiment cell volume. Some examples of clusters observed can be seen in Figure~\ref{f:clusters}. Table~\ref{t:dt_dust} lists the results obtained for each dust type and drop.

\begin{table}[b]
  \centering
  \caption{Shaking frequencies during each experiment run (in Hz). During drop 4, a glass chip blocked the shaking mechanism.}
    \begin{tabular}{|p{0.25in}||p{0.17in}|p{0.17in}|p{0.17in}|p{0.26in}|p{0.17in}|p{0.17in}|p{0.15in}|p{0.15in}|p{0.15in}|}
    \hline
    \multicolumn{1}{|p{0.25in}||}{\multirow{3}[4]{*}{\textbf{Drop}}} & \multicolumn{9}{c|}{\textbf{Flight time [s]}} \\
\cline{2-10}    \multicolumn{1}{|c||}{} & 1    & 2     & 3     & 4     & 5     & 6     & 7     & 8     & 9 \\  \cline{2-10} \cline{2-10}
	  & \multicolumn{9}{c|}{\textbf{Shaking frequency [Hz]}} \\ \hline \hline
    1     & 16.7  & 16.7  & 8.4   & 8.4   & 4.5   & 4.5   & 4.5   & 4.5   & 4.5 \\ \hline
    2     & 16.7  & 16.7  & 16.7  & 16.7/5 & 5     & 5     & 5     & 5     & 5 \\ \hline
    3     & 10.5  & 10.5  & 10.5  & 3     & 3     & 3     & 3     & 3     & 3 \\ \hline
    4     & 0   & 0   & 0   & 0   & 0   & 0   & 0   & 0   & 0 \\ \hline
    5     & 15.5  & 15.5  & 15.5  & 6     & 6     & 6     & 6     & 6     & 6 \\ \hline
    \end{tabular}
  \label{t:dt_sh_profile}
\end{table}

\section{Experiment results}
\label{s:results}

\subsection{Data analysis methods}
\label{s:analysis}

For the experiments in which cluster agglomeration was observed (see Table \ref{t:dt_dust}), aggregates and clusters could be tracked individually. In this case, the collisions in the different experiments were analysed with a semi-automatic tracking programme also used by \cite{guettler_et_al2010A&A}, \cite{weidling_et_al2012Icarus}, and \cite{kothe_et_al2013Icarus}. This programme tracks the centre-of-mass position of aggregates after frame binarisation and upon indication by the user. For each user-detected collision, it computes the absolute velocities of the colliding aggregates allowing for the determination of the relative collision velocity. In addition, the programme tracks the cross-section area of the aggregates on each two-dimensional frame. As the aggregates rotate during the time they are tracked, they present different projections to the camera and, thus, different surface areas. The average of their visible surface over their tracking time is used to calculate their mass. For each two-dimensional projection of the aggregate, its projected cross section is used to calculate an equivalent aggregate radius, i.e. the radius of a circle with the same cross section as the aggregate. The volume of the aggregate is then calculated assuming it is a sphere of its equivalent radius, and the mass is deduced assuming a constant monomer particle density (see Table~\ref{t:silicates}) and a filling factor ($\phi$=0.37). This method is fairly accurate for round aggregates and clusters and somewhat overestimates the mass for the larger fractal clusters that formed during the experiment runs.\\
In the case of no observable or only monomer agglomeration, individual aggregates were too small and too numerous to be tracked. This was the case for the spherical polydisperse dust (SP) at fast shaking frequencies during drop~1.\\

\subsection{Aggregate and cluster collisions}
\label{s:dt_collisions}

\begin{figure}[!bth]
  \begin{center}  
  \includegraphics[width=0.45\textwidth]{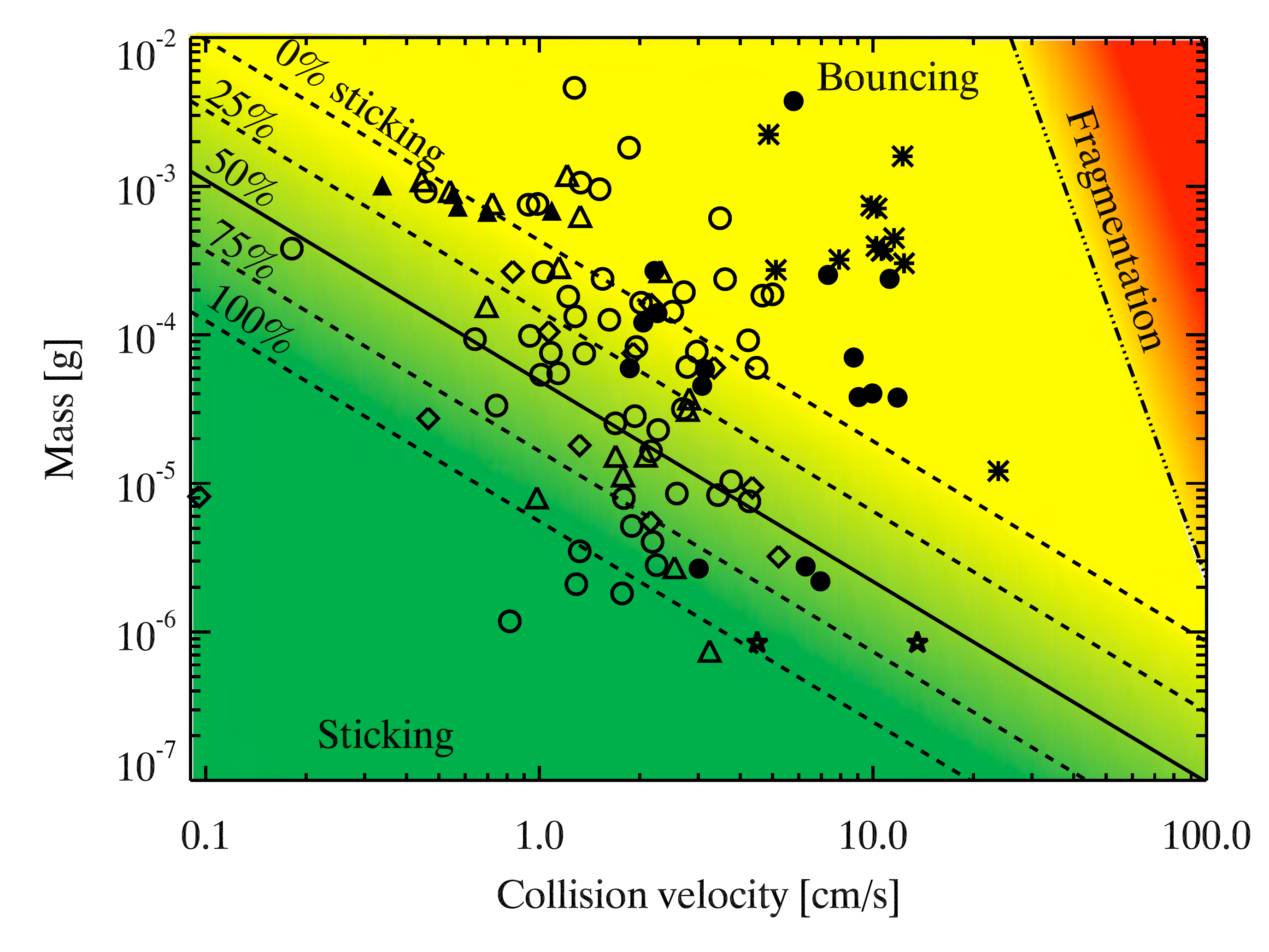}\\
 \caption{Aggregate collision outcomes observed. Open symbols indicate sticking, filled symbols indicate bouncing, and asterisks indicate fragmentation (M only). Circles denote the M sample, diamonds refer to the CM sample, triangles denote the SP sample, and stars denote the data points deduced from non-agglomerating IP. The background colours correspond to the dust collision model developed by \cite{guettler_et_al2010A&A}. Green indicates sticking, yellow indicates bouncing, and red indicates fragmentation. The lines represent the limits between sticking and bouncing (solid line for a 50\% sticking probability and dashed lines for 0, 25, 75, and 100~\% probabilities) and bouncing and fragmentation (dash-dotted line for the onset of fragmentation), which were computed by \cite{kothe_et_al2013Icarus}.}
 \label{f:res_all}
 \end{center}
\end{figure}

From the five experiment runs, data from eight cells could be analysed (see Table~\ref{t:dt_dust} for the type of result). The 162 collisions observed are presented in Figure~\ref{f:res_all}. For each collision, the mass of the smaller colliding aggregate and the collision velocity are plotted. The different symbols represent the different types of aggregates observed (see Table \ref{t:dt_dust} for dust sample codes). Open symbols represent sticking, filled symbols bouncing and asterisks mark fragmenting collisions. The M sample was the only one observed fragmenting upon mutual collision.\\
The IP sample did not cluster at all and the aggregate size distribution did not change during the experiment run. It can be assumed that all collisions that took place in this experiment cell led to bouncing of both collision partners. The mean free path of these aggregates during the experiment was calculated to be $\lambda$~=~1/($n\sigma$) =~8.31 mm, where $n$ is the number density of the aggregates inside the experiment cell and $\sigma$ the average cross section of an aggregate. This is of the same order as the experiment cell size (Figure~\ref{f:dt_cells}), and the mean aggregate collision velocity can therefore be determined as in \cite{brisset_et_al2016AA}. It is assumed that free-flying aggregates have a speed of about 1.8$v_{\textrm{max}}$, where $v_{\textrm{max}}$ is the maximum velocity of the cell walls \citep[see Figure~6 in ][where the mean collision velocity between aggregates and clusters during the SPACE suborbital flight was determined in a similar manner]{brisset_et_al2016AA}. The resulting mean collision velocities were 4.5~cm~s$^{-1}$ for drop~2 and 13.6~cm~s$^{-1}$ for drop~3. In both cases, the mean aggregate mass was 8.61$\times$10$^{-7}$ g. These two "bouncing" data points are plotted as stars in Figure~\ref{f:res_all} and represent an average over a great number of collisions.\\
The SP sample displayed two different behaviours during drop~2 and drop~5. During drop~2, the initial clusters were efficiently destroyed at a fast shaking frequency of 16.7~Hz. At the lower frequency of 5~Hz, they clustered only a few monomer aggregates at a time. These clusters were then destroyed by encounters with the cell walls before they could grow any further. They were too small and surrounded by too many free-flying aggregates to be tracked with the programme described in Section~\ref{s:analysis}. During drop~5 however, clusters already started forming during the fast shaking phase at 15.5~Hz and continued to grow at 6~Hz. In 23~sticking collisions at low velocities down to under 1~cm~s$^{-1}$, clusters grew to masses of up to about 10$^{-3}$~g, which corresponds to $\sim$1160~monomer aggregates (triangles in Figure~\ref{f:res_all}). In addition, 5~bouncing collisions at these velocity and size ranges (1~cm~s$^{-1}$ and 10$^{-3}$~g) were observed (filled triangles).\\
For the M sample, 116~collisions could be analysed. The outcome of these collisions was sticking for 88~of them, bouncing for 16~and fragmentation for~12. With relative velocities down to a few~mm~s$^{-1}$, these aggregates formed clusters of up to about 10$^{-2}$~g, which corresponds to $\sim$11600 monomer aggregates. The bouncing collisions were observed at higher velocities between 2~and~11~cm~s$^{-1}$ for collision partner sizes ranging from the initially prepared aggregate ($\sim10^{-6}$ g) to the bigger clusters formed during the experiment run ($\sim10^{-3}$~g). At about 5~cm~s$^{-1}$, the first fragmenting collision was observed. Except for one, which took place at the highest observed velocity of 23.8~cm~s$^{-1}$, the other fragmenting collisions happened in collisions between bigger clusters (2$\times10^{-4}$~g and more).\\
In the CM sample, we observed 18~collisions resulting in sticking (diamonds). Neither bouncing nor fragmenting collisions between these aggregates could be detected. The aggregates were observed during drop~3, which had the slowest shaking profile from all considered drops (10.5~and 3~Hz shaking frequencies compared to~$>$15 and $>$5~Hz for drops 2~and~5). Therefore, the collision velocities were also smaller, ranging from 0.1~to~6~cm~s$^{-1}$. The corresponding calculated masses of these clusters were 5$\times10^{-4}$~g, which would be more than one order of magnitude smaller than for the M sample, if the volume filling factors were the same for both.\\

\subsection{Cluster restructuring and fragmentation}
\label{s:wall_collisions}

\begin{figure}[t]
  \begin{center}  
  \includegraphics[width=0.5\textwidth]{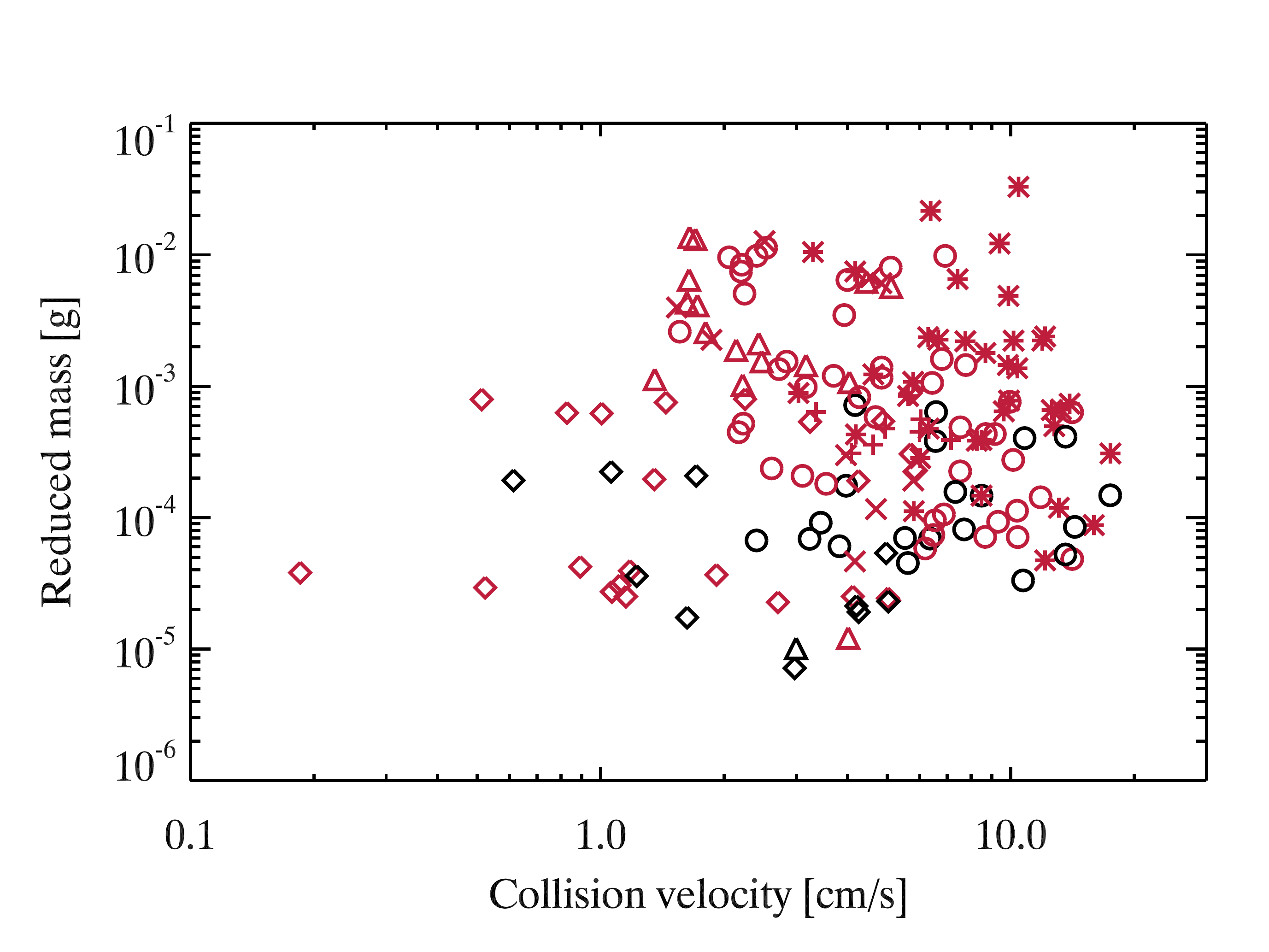}\\
 \caption{Outcome of collisions between aggregates or clusters with the experiment cell walls. Only bouncing and fragmentation collisions were observed. Circles and asterisks indicate the M sample, diamonds, and plus signs indicate the CM sample, and triangles and x signs indicate the SP sample. Collisions for which aggregate or cluster restructuring was visible are shown in red.}
 \label{f:dt_col_wall}
 \end{center}
\end{figure}
\begin{figure}[t]
  \begin{center}  
  \includegraphics[width=0.5\textwidth]{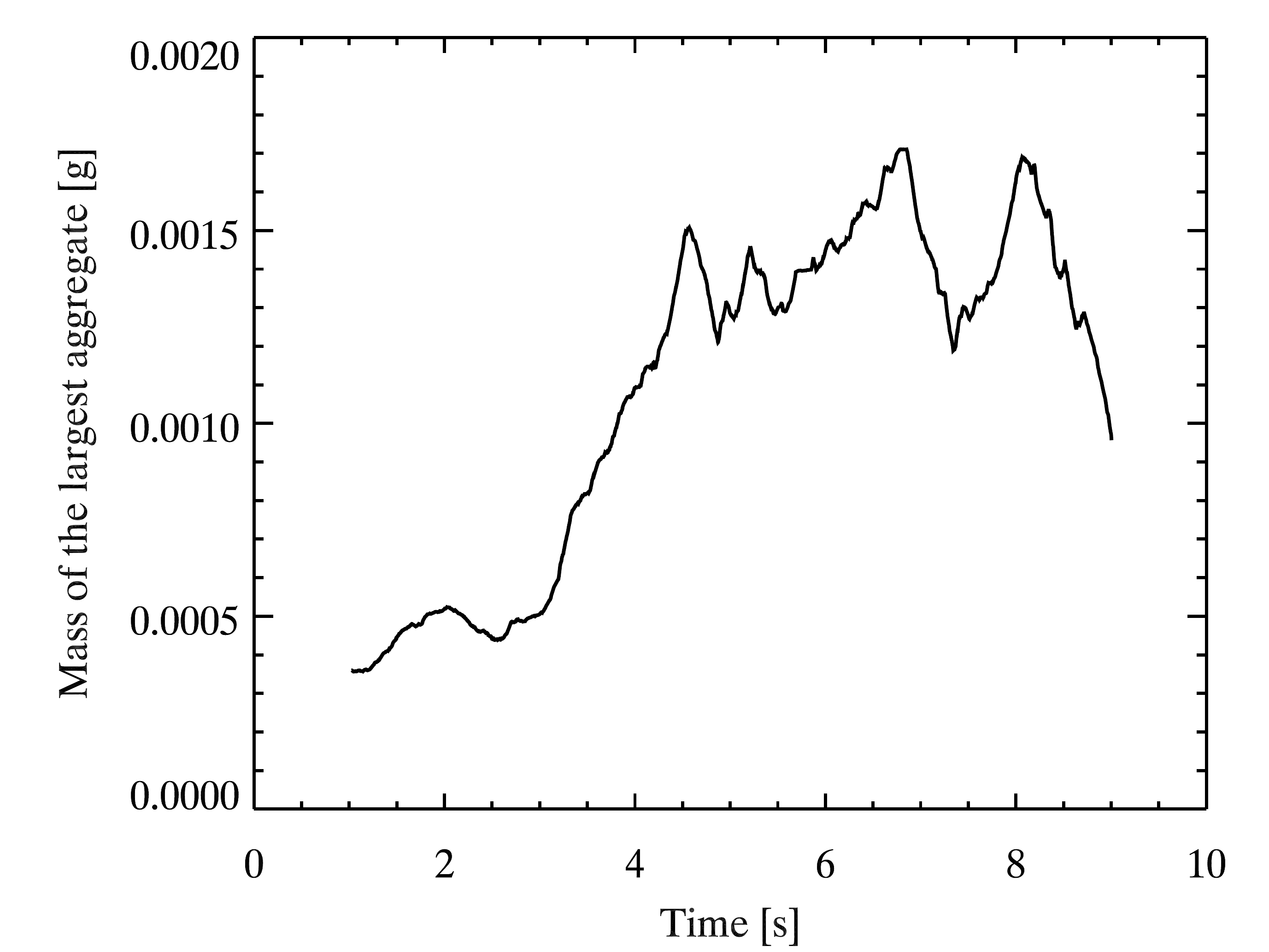}\\
 \caption{Size evolution of the largest cluster in cell 3 during drop 5 (M sample).}
 \label{f:agg_sizes}
 \end{center}
\end{figure}

In addition to the collisions between aggregates and clusters, the data also showed collisions of clusters with the experiment cell walls. These clusters were tracked and the collision properties were deduced (Figure~\ref{f:dt_col_wall}). The highest observed collision velocities were around the maximum wall speed of $\sim$20~cm~s$^{-1}$, corresponding to the maximum shaking frequency of 16.7~Hz. The mass presented is the reduced mass of the collision, i.e. the mass of the cluster, as in comparison, the wall is considered to have an infinite mass.\\
For the M sample, 64~bouncing and 37~fragmenting cluster-wall collisions were observed. For smaller clusters of masses between 10$^{-4}$ and 10$^{-3}$~g, both collision outcomes were observed for the maximum velocities of $\sim$20~cm~s$^{-1}$. Bigger clusters, however, fragmented more frequently than they bounced. The two other types of aggregates (CM and SP) also displayed a transition from bouncing to fragmentation with increasing cluster mass and velocity after cluster collisions with the cell walls. For the CM sample, 34~bouncing and 8~fragmenting cluster wall collisions were observed. For the SP sample, 17~bouncing and 9~fragmenting cluster wall collisions were observed.\\
Figure~\ref{f:dt_col_wall} shows all the collisions between clusters and the cell wall, showing the reduced mass of the collision and collision velocity. Collisions where cluster restructuring was visible are plotted in red. For the M and SP samples, most of the restructuring events occurred at velocities and masses higher than 5~cm~s$^{-1}$ and 10$^{-4}$~g, respectively, and became more frequent with increasing velocity and cluster mass. For the CM sample, cluster restructuring events already occurred at velocities as low as 0.19~cm~s$^{-1}$ for masses around 4$\times10^{-5}$~g.  \\
Even at higher relative velocities, the bigger clusters observed colliding with the cell walls had masses of up to $\sim10^{-3}$~g. The SP clusters formed during the fast phase of the shaking profile of the experiment in a growth process just like the bigger clusters observed colliding with each other inside of the cell volume (see Section~\ref{s:dt_collisions}). In a different manner, M and CM clusters were remnants of the drop tower capsule launch. They formed while aggregates were lying in a heap on the bottom of the cell and were pressed together when the capsule was accelerated. Afterwards, they were partially destroyed by the fast shaking at the beginning of the microgravity phase, which reduced their size. Figure~\ref{f:agg_sizes} shows the size evolution of the largest cluster in cell 3 (M sample) during drop~5. While a clear increase in growth rate can be seen as soon as the shaking frequency is reduced (3~s; see Table~\ref{t:dt_sh_profile}), the largest aggregate in the cell is already a cluster of $\sim$10$^{-4}$~g during the fast shaking phase. It can be noted that dips in the size of the largest aggregate are introduced by fragmenting collisions with the cell walls.\\

\section{Inputs to the dust collision model}
\label{s:model_input}

The results for collisions between aggregates obtained in the different experiment cells are summarised in Figure~\ref{f:res_all}. Collisions are plotted as single points and the mass given is the mass of the smaller collision partner. The dust collision model developed by \cite{guettler_et_al2010A&A} and \cite{kothe_et_al2013Icarus} is shown in the background, delimiting the regions of the parameter field where sticking (green), bouncing (yellow), and fragmentation (red) are expected for same-sized dust aggregate collisions.

\subsection{The sticking to bouncing transition}
\label{s:st-b}
\begin{figure}[t]
  \begin{center}  
  \includegraphics[width=0.45\textwidth]{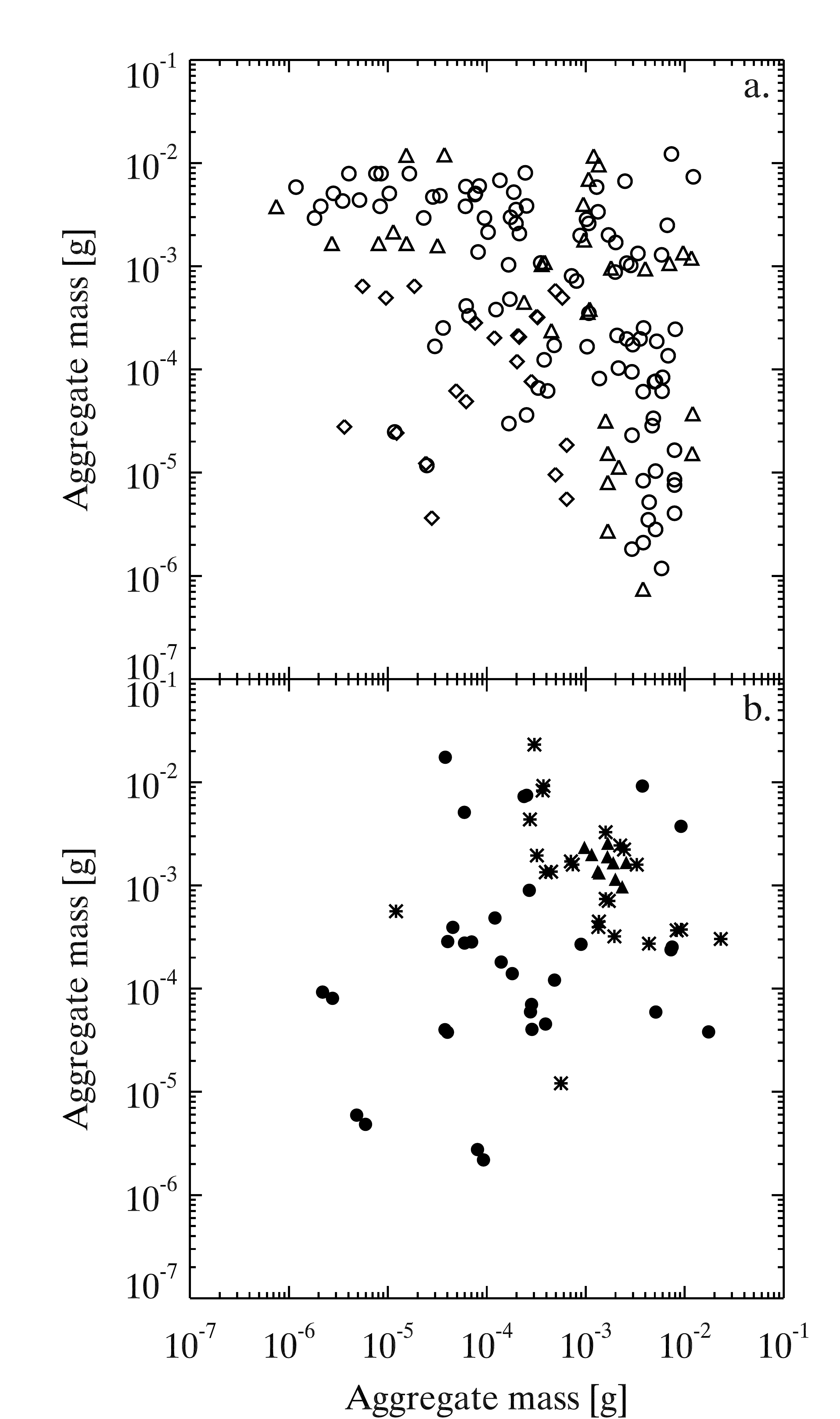}\\
 \caption{Masses of aggregates involved in collisions observed: a. sticking collisions (open symbols). b. bouncing (filled symbols) and fragmenting (asterisks) collisions. Circles: M, diamonds: CM, triangles: SP. In both plots, each collision is represented twice (symmetrically with respect to the diagonal). The sticking collisions were separated from the others for better legibility.}
 \label{f:butterfly}
 \end{center}
\end{figure}

An overall tendency can be recognised in the gathered data: the likelihood of sticking decreases with increasing cluster size and collision velocity. The mean relative velocity and reduced aggregate mass for sticking collisions are 2.1$\pm0.9$~cm~s$^{-1}$ and 2.2$^{+16}_{-2.1}\times10^{-4}$~g, respectively, in contrast to 5.9$\pm$3.2~cm~s$^{-1}$ and 7.8$\pm$6.3$\times10^{-5}$~g for the bouncing collisions (see Table~\ref{t:com_aggr}, M sample). The coexistence of sticking and bouncing collisions in a region of the parameter field along the transition line has also been observed by \citet{weidling_et_al2012Icarus} and \cite{kothe_et_al2013Icarus}. \\
However, it is obvious that many sticking collisions were taking place in parameter ranges where bouncing would be expected in the current model. For example, at the same mean aggregate sizes of 2.2$\times10^{-4}$~g, \cite{guettler_et_al2010A&A} extrapolated a 50\% sticking to bouncing transition velocity of 7.2$\times10^{-2}$~cm~s$^{-1}$. For the same aggregate mass, \cite{kothe_et_al2013Icarus} computed a transition velocity of 0.33~cm~s$^{-1}$. This sticking at higher collision velocities can be explained by the fact that most of the observed events were not aggregate-aggregate collisions, for which the model was developed, but aggregate-cluster or cluster-cluster collisions. To illustrate this, Figure \ref{f:butterfly} plots both aggregate/cluster masses for each collision observed, for sticking collisions (a) and bouncing and fragmenting collisions (b). Most of the sticking collisions observed did not involve monomer aggregates ($\sim$8$\times10^{-7}$~g), but clusters of $>$10$^{-4}$~g in mass, composed of more than 100 monomer aggregates. The enhanced sticking probability of clusters composed of a high number of aggregates was also observed by \cite{kothe_et_al2013Icarus}, compared with the results of \citet{weidling_et_al2012Icarus} who analysed aggregates of 10$^{-4}$~g to 10$^{-3}$~g in mass that resulted in bouncing for more than 90~\% of the collisions at velocities between about 0.2~and~50~cm~s$^{-1}$.\\
Figure \ref{f:butterfly} also shows that the mass ratio between colliding aggregates and clusters in the different experiment runs covers up to 4 orders of magnitude. At the highest mass ratio of $\sim$10$^4$, only sticking collisions were observed. The highest mass ratio between aggregates and clusters involved in a bouncing or fragmenting collision is about 10$^2$.\\

\begin{table}[tph]
  \centering
  \caption{Mean relative velocity and reduced mass for collisions in the M sample, with respect to their collision outcomes.}
    \begin{tabular}{|l|>{\centering\arraybackslash}p{1in}|>{\centering\arraybackslash}p{0.85in}|}
    \hline
    \textbf{Collision outcome} & \textbf{Relative velocity (cm~s$^{-1}$)} & \textbf{Reduced mass (g)}\\ \hline
    sticking& 2.1$\pm0.9$ & 2.2$^{+16}_{-2.1}\times10^{-4}$ \\  \hline
   bouncing  & 5.9$\pm$3.2 & 7.8$\pm$6.3$\times10^{-5}$ \\  \hline
    fragmentation&10.1$\pm$3.2& 4.2$\pm$2.4$\times10^{-4}$\\\hline
    \end{tabular}
  \label{t:com_aggr}
\end{table}

\subsection{The bouncing to fragmentation transition}

\begin{figure}[t]
  \begin{center}  
  \includegraphics[width = 0.45\textwidth]{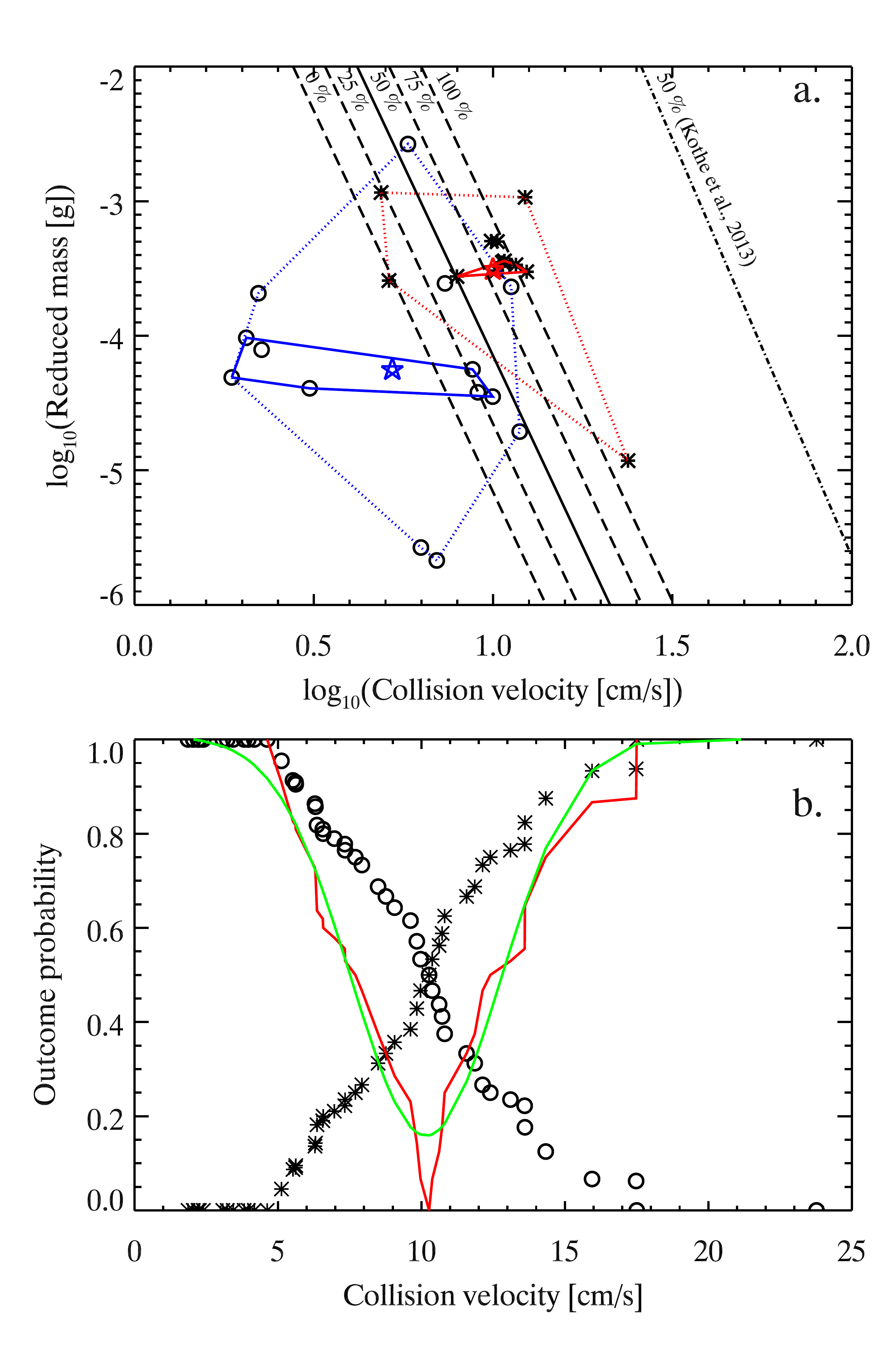}\\
 \caption{a. Transition between bouncing (circles, blue)  and fragmentation (asterisks, red) for M clusters. The mean of each set of data points are denoted by the stars. The solid contours enclose half of the respective data points around their mean value (50~\% occurrence) and the dotted contours enclose all of the data points (100~\% occurrence). The solid black line indicates the computed 50~\% transition between bouncing and fragmentation, and the dashed lines the corresponding 0, 25, 75, and 100~\% probability for fragmentation (derived from the bouncing and fragmentation probabilities shown in b.). The dash-dotted line on the right side of the plot represents the fragmentation onset in the collision model of Kothe, et al. (2013). b. Bouncing (open circles) and fragmentation (asterisks) probabilities for collisions between clusters composed of aggregates of monodisperse dust. The red curve shows the difference in outcome probability and the green curve shows its Gaussian fit. This probability difference was used to compute the 0, 25, 75, and 100~\% probability levels shown as dashed lines in a. (see details in text).}
 \label{f:com_aggr}
 \end{center}
\end{figure}

As can be seen in Figure~\ref{f:res_all} (asterisks), the fragmenting collisions (observed only between clusters of non-compacted aggregates composed of monodisperse SiO$_2$) happened at masses and velocities where bouncing would be expected in the dust collision model \citep{guettler_et_al2010A&A,kothe_et_al2013Icarus}. In the M sample, collisions leading to bouncing had a mean collision velocity of 5.9$\pm$3.2~cm~s$^{-1}$ and a reduced mass of 7.8$\pm$6.3$\times10^{-5}$~g. Collisions leading to fragmentation of the cluster had a mean collision velocity and reduced mass of 10.1$\pm$3.2~cm~s$^{-1}$ and 4.2$\pm$2.4$\times10^{-4}$~g, respectively (Table~\ref{t:com_aggr} and stars in Figure~\ref{f:com_aggr}). These values confirm the general trend of cluster collisions to transit from bouncing to fragmentation with growing collision velocities. Figure~\ref{f:com_aggr} plots the bouncing (circles) and fragmentation (asterisks) collisions for M clusters according to their collision velocity and reduced mass, together with the respective mean values (stars). Also shown are the contours of 50~and~100~\% occurrence of bouncing and fragmentation. These contours enclose half (50~\% occurrence) and all (100~\% occurrence) of the respective data points, centred around the mean of each group of points.\\
A transition between bouncing and fragmentation was computed with the method presented in \cite{kothe_et_al2013Icarus}. For this transition, a power law relation was assumed, of the form
\begin{equation}
\frac{m}{1\,\textrm{g}} = 10^{a}\left(\frac{v}{1\,\textrm{cm~s$^{-1}$}}\right)^{b}
\end{equation}
where $m$ is the mass of the smaller cluster, $v$ the relative collision velocity and $a$, $b$ the fit parameters. Three fitting methods were applied to the available set of bouncing and fragmentation data points: least squares, least linear and least number of data points deviation, with both asymmetric and symmetric false data point discrimination. This resulted in six fits optimised by different criteria. The method of conditional value at risk \citep[CoVar,][]{hull2012} was used to choose the best fit out of these six sets of values \citep[see][for details]{kothe_et_al2013Icarus}. The best fit was reached for the least squares deviation method with an asymmetric false data point discrimination, and the parameters $a$~=~1.52 and $b$~=~-5.67. This fit is also represented in Figure \ref{f:com_aggr} (solid black line). The bouncing and fragmentation probabilities computed at a given collision speed are shown in Figure~\ref{f:com_aggr}b. By taking the difference between them at each (binned) collision speed, we can get a measure for the transition from bouncing to fragmentation: the collision speed for which fragmentation becomes more probable than bouncing is considered to be the 50\% transition speed. The Gaussian fit to the probability difference gives the width of the transition, from 0 to 100\% bouncing. The 0, 25, 75 and 100\% fragmentation probability limits are also plotted in Figure~\ref{f:com_aggr}a (dashed lines).\\
The computed transition has an offset of a factor of 6 in velocity compared with the transition line computed by \cite{kothe_et_al2013Icarus} (dash-dotted line on the right side of Figure~\ref{f:com_aggr}a). This difference can be attributed to the nature of the collision partners. To compute the transition between bouncing and fragmentation, \cite{kothe_et_al2013Icarus} used collision data gathered by \cite{blum_and_muench1993Icarus, beitz_et_al2011ApJ,deckers_and_teiser2013ApJ}, and \cite{schraepler_et_al2013EPSC}. All of these experiments were performed with aggregates prepared in the laboratory that had a three-dimensional fractal dimension of 3. The collision partners that were seen to be fragmenting, however, were clusters built during the microgravity phase of the experiment that had a mean fractal dimensions lower than 3. This lower fractal dimension made them more fragile than the aggregates prepared in the laboratory.

\section{Discussion}
\label{s:discussion}

\subsection{100 $\mu$m-sized aggregates as building blocks for cluster growth}
\label{s:size_relevance}

Micrometre-sized particles are usually considered building blocks for the growth of aggregates and clusters inside PPDs. This is mostly motivated by the observations of $\mu$m-sized dust grains in discs around young stars \citep{dalessio_et_al2001ApJ,vanboekel_et_al2003AA,testi_et_al2014PP} and the hit-and-stick behaviour attributed to these grain sizes \citep{johansen_et_al2014PP,testi_et_al2014PP,blum_wurm2008ARAA}. However, recent experimental and numerical work has demonstrated that the growth process of dust grains cannot proceed through simple sticking collisions to planetesimals sizes, for which gravitational forces take over \citep{zsom_et_al2010AA}. Solutions to overcome growth impediments, such as the bouncing or metre barriers, might include the drift and local concentration of dust grains, leading to the formation of planetesimals directly from $\sim$1 to 10 mm-sized building blocks \citep{yang2016}. When the conditions in the PPD are such that only sub-mm-sized aggregates can form before they enter the bouncing regime, drift into calmer regions might render those aggregates into building blocks of clusters grown by further sticking collisions. This work was dedicated to the study of the conditions under which this cluster formation is possible.

\subsection{Collision recipe for clusters composed of sub-mm-sized aggregates}
\label{s:recipe}

The collisions between clusters and the cell walls led to bouncing, with or without visible restructuring, or to fragmentation of the clusters (Figure~\ref{f:dt_col_wall}). This set of collisions with the cell walls was augmented with additional collisions observed between these same aggregates or clusters (30 collisions for M and 10 for CM), including sticking events where restructuring of the cluster was observed. The clusters were composed of aggregates of about 120~$\mu$m in diameter, but displayed a collision behaviour very similar to those in molecular dynamics simulations with the clusters composed of sub-$\mu$m-sized monomer particles \citep[e.g.][]{dominik_and_tielens1997ApJ,wada_et_al2009ApJ}. As mentioned in \cite{brisset_et_al2016AA}, where the aggregate surface energy was measured for $\sim$100~$\mu$m-sized aggregates, the possibility to scale the surface energy of $\mu$m-sized particles to macroscopic ($\sim$100~$\mu$m) aggregates indicates that these collision behaviours simulated with microscopic particles ($\sim1\mu$m) could be extended to clusters composed of macroscopic aggregates.\\
The outcome of simulated collisions between clusters composed of $\mu$m-sized solid particles follows a collision "recipe" introduced by \cite{dominik_and_tielens1997ApJ}. This recipe defines three threshold energies $E_{\textrm{stick}}$, $E_{\textrm{roll}}$ and $E_{\textrm{break}}$ and can be summarised in terms of collision energy as follows \citep[see Table 3 of][]{dominik_and_tielens1997ApJ}: 
\begin{itemize}
\item up until a certain collision energy $E_{\textrm{stick}}$, monomer aggregates always stick to the target cluster they collide with,
\item for a collision energy of 5$E_{\textrm{roll}}$, cluster restructuring becomes visible,
\item for a collision energy of 3$N_{\textrm{c}}E_{\textrm{break}}$, clusters start to lose monomer aggregates, where $N_{\textrm{c}}$ is the number of monomer-monomer contacts inside the cluster, 
\item for a collision energy of 10$N_{\textrm{c}}E_{\textrm{break}}$, clusters are disrupted completely.
\end{itemize}
\begin{table*}[bhpt]
  \centering
  \caption{Microscopic and macroscopic threshold energies for cluster restructuring and fragmentation measured for monodisperse dust.}
    \begin{tabular}{|>{\centering}p{0.7in}|>{\centering}p{1.5in}|>{\centering}p{1.5in}|>{\centering\arraybackslash}p{1.5in}|}
    \hline
    \textbf{Threshold energy} & \textbf{Microscopic value [J]} & \textbf{Macroscopic value for M clusters [J]} & \textbf{Macroscopic value for CM clusters [J]}\\  \hline
    $E_{\textrm{roll}}$& 1.7$\times10^{-15}$ & 1.8$\pm0.9\times10^{-13}$ & 5.8$\pm4.2\times10^{-15}$\\
    			& \citep{heim_et_al1999PRL} & &\\ \hline
    $E_{\textrm{break}}$& 1.3$\times10^{-15}$ & 3.5$\pm1.5\times10^{-13}$ & 3.3$\pm1.4\times10^{-13}$\\
     			& \citep{poppe_et_al1999AdSR} && \\ \hline
    \end{tabular}
  \label{t:en_macro}
\end{table*}
The parameter $E_{\textrm{roll}}$ is the critical rolling energy defined as the energy required to start irreversible rolling of one monomer particle over another. The parameter $E_{\textrm{break}}$ is the critical breaking energy and is defined as the energy required to break a contact between two monomer particles. The number of contacts $N_{\textrm{c}}$ is determined by the coordination number of the aggregates inside a cluster. The clusters in the cell volume grew either through cluster-cluster or through aggregate-cluster collisions, which would both have led to very low average coordination numbers of less than~2. However, as they experienced frequent collisions with the cell walls leading to compaction, the choice of a higher coordination number is necessary; random closed packing has a coordination number of~8.4. Clusters generated by ballistic agglomeration with migration \citep[BAM2; see][]{shen_et_al2008ApJ} have a volume filling factor of~$\sim$0.4 for a coordination number of~6, which seems appropriate for the clusters studied here, as they grew through aggregate-aggregate collisions and then compacted through collisions with the cell walls. The fractal dimension is chosen to be D$_{\textrm{f}}$ = 1.80 (three-dimensional fractal dimension estimated from the two-dimensional fractal dimension using the method described in \cite{kothe_et_al2013Icarus}). Accordingly, the total number of contacts between aggregates of radius $r_0$ in a cluster of radius $r$ is
\begin{equation}
N_{\textrm{c}} = 6\left(\frac{r}{r_0}\right)^{1.80}
\label{e:nc}
\end{equation}
\begin{center}
\begin{figure}[t]
       \includegraphics[width=0.5\textwidth]{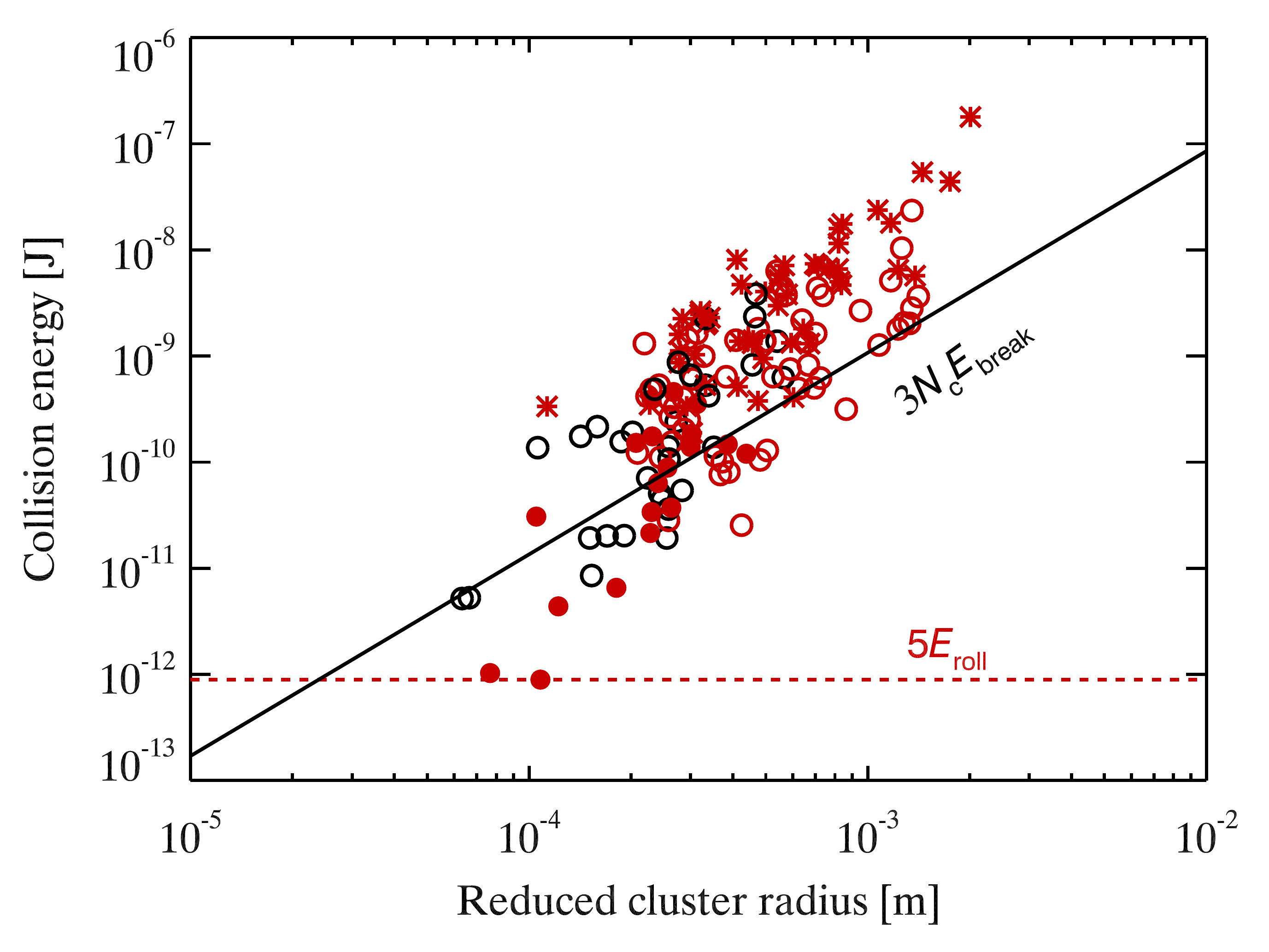}
 	\caption{Cluster collisions, including collisions with the experiment cell walls for the M sample. Filled circles indicate sticking, open circles indicate bouncing, and asterisks indicate fragmentation. Red points show the collisions that displayed visible restructuring. The threshold energies 5$E_{roll}$ for onset of visible restructuring (red dashed line) and 3N$_cE_{break}$ for the loss of the first monomers (solid) are represented.}
	 \label{f:recipe}
\end{figure}
      \end{center}
      \begin{center}
      \begin{figure}[t]
      	\includegraphics[width=0.5\textwidth]{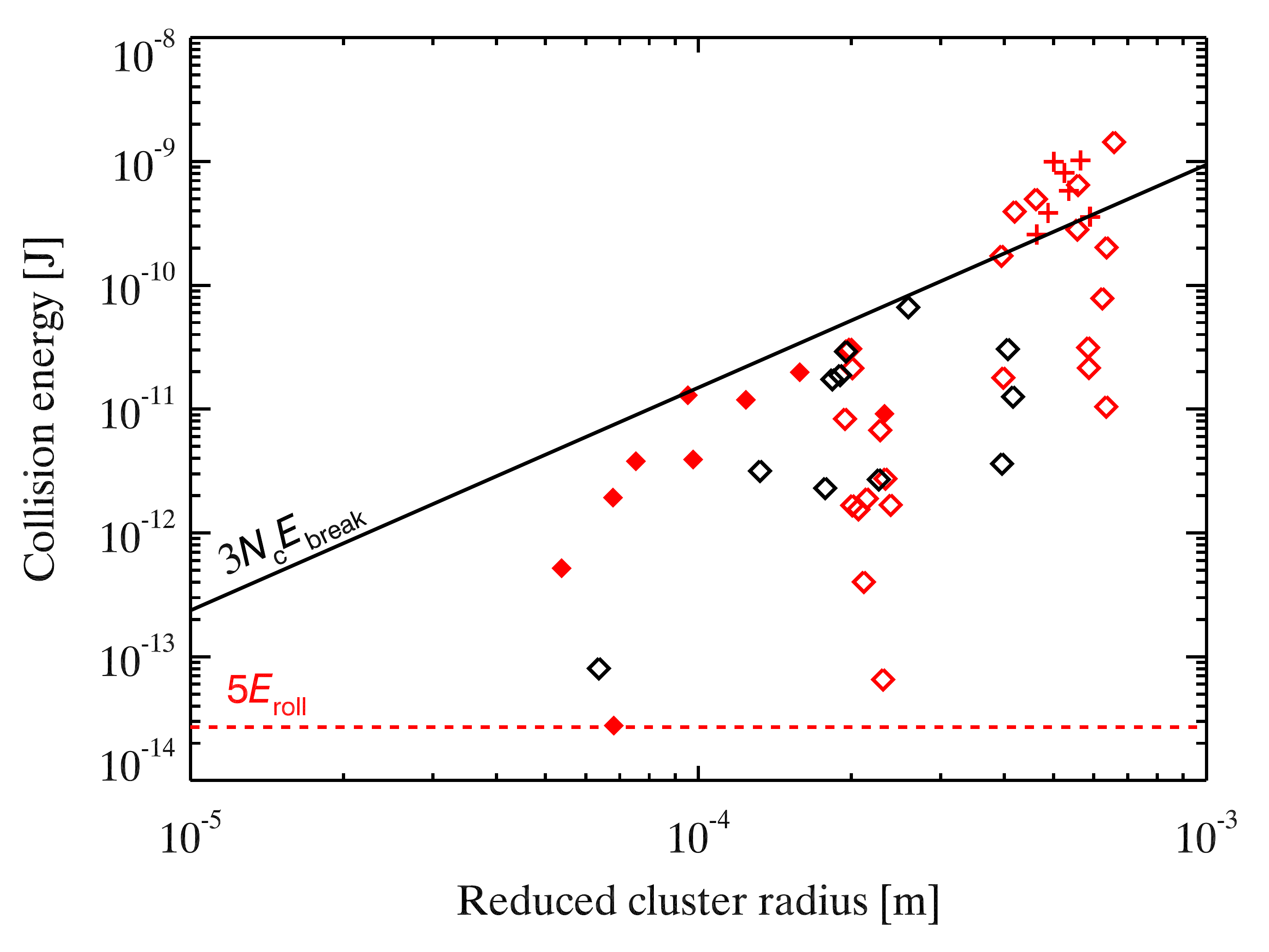}\
	 \caption{Same as Figure~\ref{f:recipe} for the CM sample. Filled diamonds indicate sticking, open diamonds indicate bouncing, and plus signs indicate fragmenting. Red points show the collisions that displayed visible restructuring.}
      \label{f:recipe_c}
\end{figure}
\end{center}
In the data gathered, cluster restructuring and fragmentation could be observed directly. Figure~\ref{f:recipe} shows the collisions of clusters either with individual monomer aggregates or with the cell walls, according to the collision energy and reduced mass of the collision, for the M sample. The collision energy E$_{\textrm{coll}}$ was approximated to be composed of the translational kinetic energy of the colliding particle system only,
\begin{equation}
E_{\textrm{coll}}=\frac{1}{2}m_{\textrm{red}}v_{\textrm{rel}}^2,
\end{equation}
where $m_{\textrm{red}}$ is the reduced mass of the collision and $v_{\textrm{rel}}$ the relative velocity between the cluster and the cell wall;  in the case
of a collision with the wall, the reduced mass is the mass of the cluster,
as the mass of the wall is considered infinitely large. The reduced radius and reduced mass were derived from the area measurement of the aggregates/clusters during the experiment run (see Section~\ref{s:analysis}). Sticking collisions are indicated as filled circles, bouncing collisions as open circles, and fragmenting collisions (loss of monomers) as asterisks. Red data points indicate that restructuring of the cluster was visible. Sticking collisions were observed at collision energies ranging from about $10^{-12}$ to 5~$10^{-10}$~J and radii ranging from 0.07 to 0.5~mm. Bouncing collisions with the cell walls were observed at collision energies between $10^{-11}$~and~$10^{-8}$~J for radii between 0.1~and~1~mm, while fragmenting collisions were seen at energies between $10^{-10}$~and~$10^{-7}$~J for radii between 0.1~and~2~mm. The horizontal red dashed line represents the minimum energy at which visible restructuring of the clusters was observed: 5$E_{\textrm{roll}}=8.9\times10^{-13}\pm7.7\times10^{-16}$~J. From this value a macroscopic $E_{\textrm{roll}}$ was calculated: $E_{\textrm{roll}} = 1.8\pm0.9\times10^{-13}$~J.\\
In the same manner a value for the energy required to break a contact between two macroscopic aggregates was derived: the solid line in Figure~\ref{f:recipe} indicates the onset energy of fragmentation, 3$N_{\textrm{c}}E_{\textrm{break}}$, with $N_{\textrm{c}}$ defined in Equation~\ref{e:nc}. The measured value was $E_{\textrm{break}}$ = 3.5$\pm1.5\times10^{-13}$~J.\\
The same analysis was performed on the CM sample (Figure~\ref{f:recipe_c}). Table~\ref{t:en_macro} compares the rolling and breaking energies for the compacted and non-compacted aggregates. The aggregate compaction seems to have no significant influence on the energy required to break a contact between two aggregates ($E_{\textrm{break}} = 3.3\times10^{-13}$~J and 3.5$\times10^{-13}$~J for M and CM aggregates, respectively). The rolling energy, however, is 2~orders of magnitude lower for the CM (5.8$\times10^{-15}$~J) than for the M aggregates (1.8$\times10^{-13}$~J), indicating that irreversible rolling is much easier to trigger between aggregates with a compacted shell than between uniformly porous aggregates.

\subsection{SPACE results application to protoplanetary discs}
\label{s:ppd_application}

The collisions observed during the different experiment runs occurred between free-floating aggregates. The semi-automatic tracking of the particles revealed that the walls were inert in reflecting the aggregates, thus only contributing to the velocity distribution of the aggregates. The aggregates tracked to obtain the data represented in Figure~\ref{f:res_all} did not encounter any of the cell walls during the time they were tracked (five\ frames minimum). \\
Our collision results can therefore be compared with current nebula models so that conclusions on the sub-mm-sized dust aggregate behaviour in protoplanetary discs can be drawn. The three solar nebula models -- the Minimum Mass Solar Nebula (MMSN) model \citep{weidenschilling1977ASS}, the low-density model \citep{andrews_and_williams2007ApJ}, and the high-density model \citep{desch2007ApJ} -- considered are described in detail in Section~4.4 of \cite{brisset_et_al2016AA}.  

\begin{figure}[ht]
  \begin{center}  
  \includegraphics[width=0.39\textwidth]{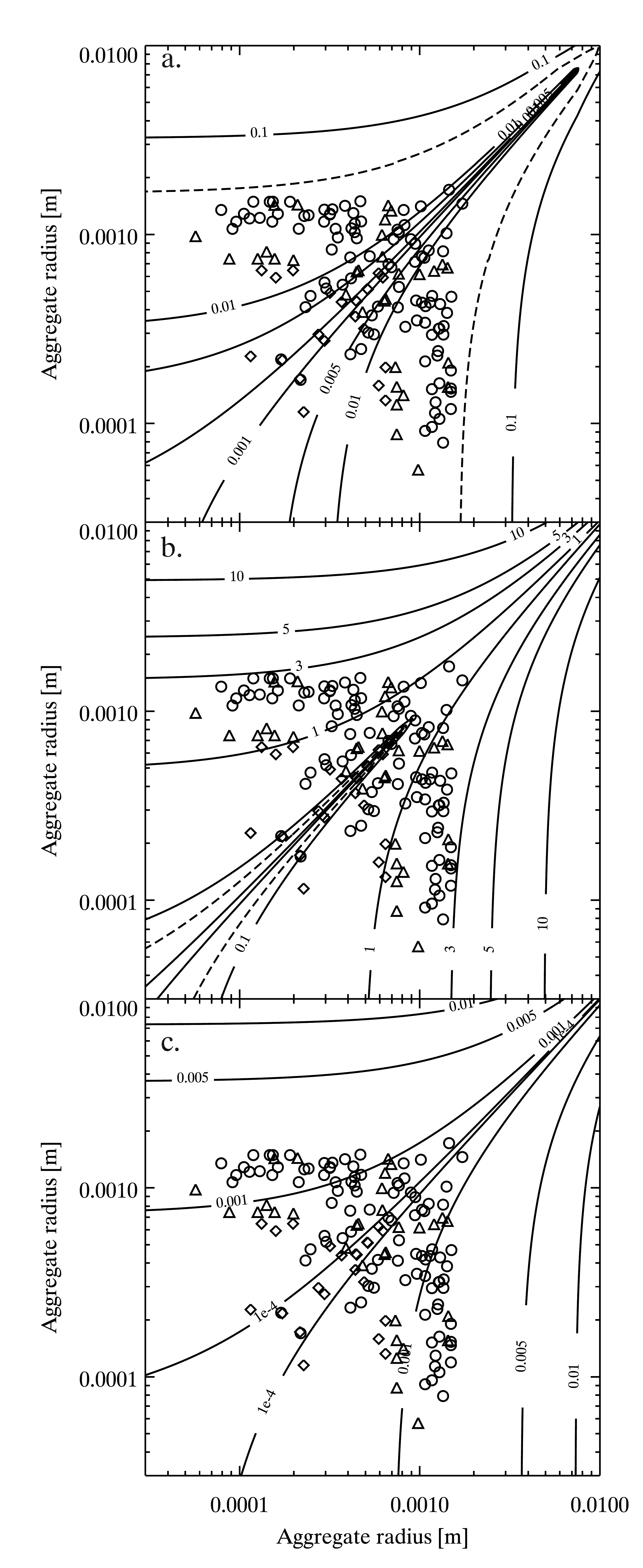}\\
 \caption{Relative velocities between dust aggregates in a protoplanetary disc computed according to panel a, \citet{weidenschilling1977ASS}; panel b, \citet{andrews_and_williams2007ApJ}; and panel c. \citet{desch2007ApJ} at 1~AU and a turbulence parameter of $\alpha = 10^{-5}$. The velocity profiles are labelled in units of m~s$^{-1}$. The sticking collisions observed are plotted as single points; circles indicate the M sample, diamonds indicate the CM sample, and triangles indicate the SP sample. The maximum speed at which sticking was observed (5.2 cm~s$^{-1}$)is shown as a dashed line. In panel c, the dashed contour is outside of the depicted parameter field.}
 \label{f:hay}
 \end{center}
\end{figure}

\paragraph{\textit{Aggregate sticking and growth}}

The computed relative velocity profiles for a low turbulence region (dead zone, $\alpha$=10$^{-5}$) at 1~AU can be seen in Figure~\ref{f:hay} for all three models \citep{weidenschilling1977ASS,andrews_and_williams2007ApJ,desch2007ApJ}. The collisions observed during the different experiment runs are plotted as well. From this figure, the relative velocities between the colliding aggregates that would be induced by a PPD environment can be derived. In order to compare these relative velocities to those induced by the experimental set-up, the maximum velocity at which sticking was observed, 5.2~cm~s$^{-1}$ (see Section~\ref{s:dt_collisions}) is shown as a dashed contour in Figure~\ref{f:hay}. \\
For the MMSN (Figure~\ref{f:hay}a.) and the compact (Figure~\ref{f:hay}c.) nebula models, the expected relative velocities for aggregate-aggregate and aggregate-cluster collisions of these sizes are lower than the highest observed velocity measured for sticking. This indicates that at 1~AU and in a low turbulence environment such as an MRI dead zone, the investigated collisions would lead to cluster growth. In fact, collisions would lead to cluster growth up to sizes of a few~mm for the Minimum Mass Solar Nebula \citep{weidenschilling1977ASS} and for the compact model \citep{desch2007ApJ}. \\


\paragraph{\textit{Cluster restructuring and fragmentation}}

\begin{figure}[t]
  \begin{center}  
  \includegraphics[width = 0.43\textwidth]{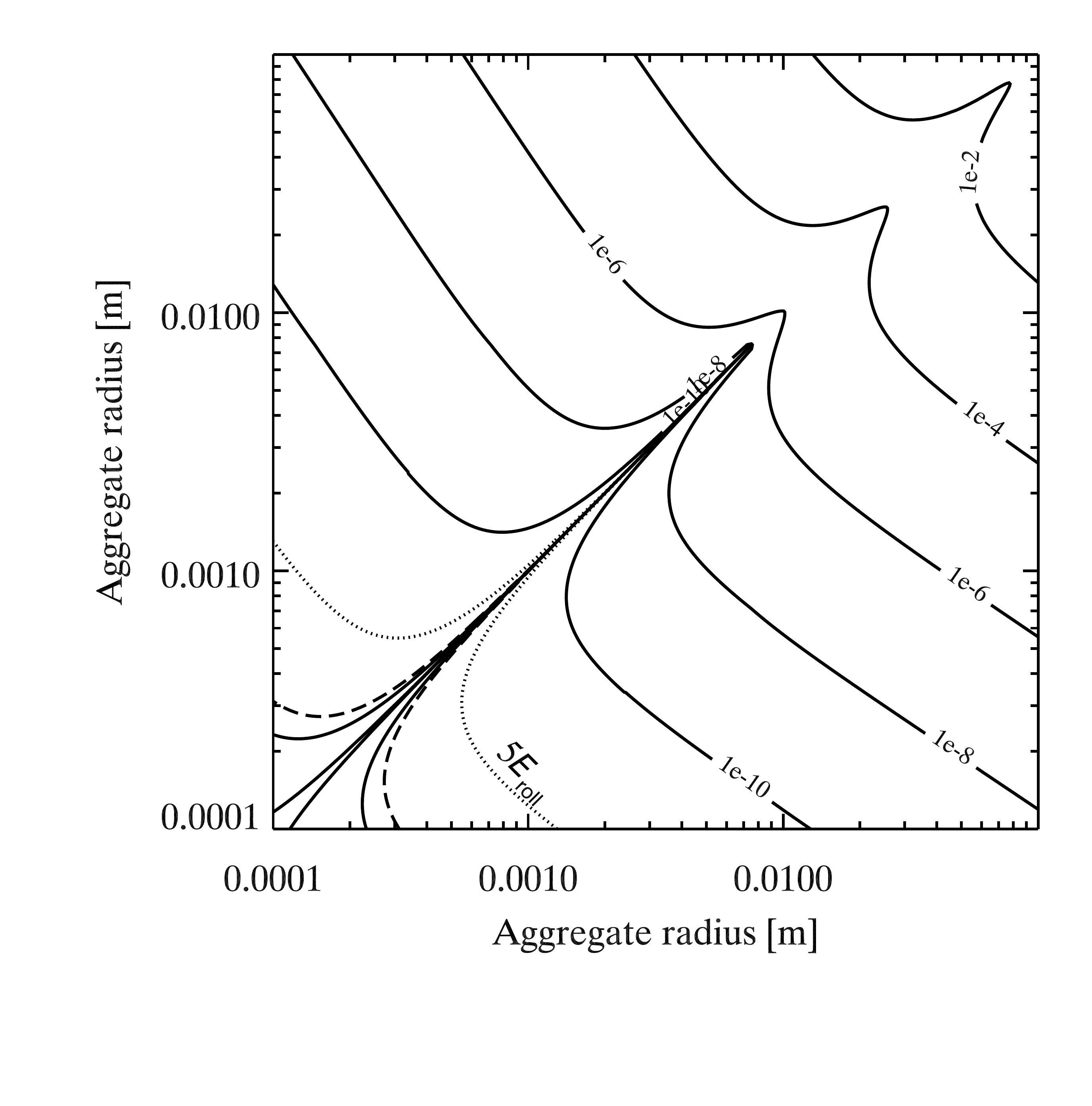}
 \caption{Collision energy profiles (in J) for clusters composed of $\sim$100 $\mu$m aggregates in a MMSN model at 1 AU with $\alpha = 10^{-5}$. The dotted contour represents the minimal energy for onset of cluster restructuring $5E_{\textrm{roll}}$~=~8.9$\times10^{-13}$ J measured in Section~\ref{s:recipe} for non-compacted aggregates. The value for compacted aggregates (2.7$\times10^{-14}$ J) is shown as a dashed contour.}
 \label{f:hay_en}
 \end{center}
\end{figure}

\begin{figure}[t]
  \begin{center}  
  \includegraphics[width = 0.43\textwidth]{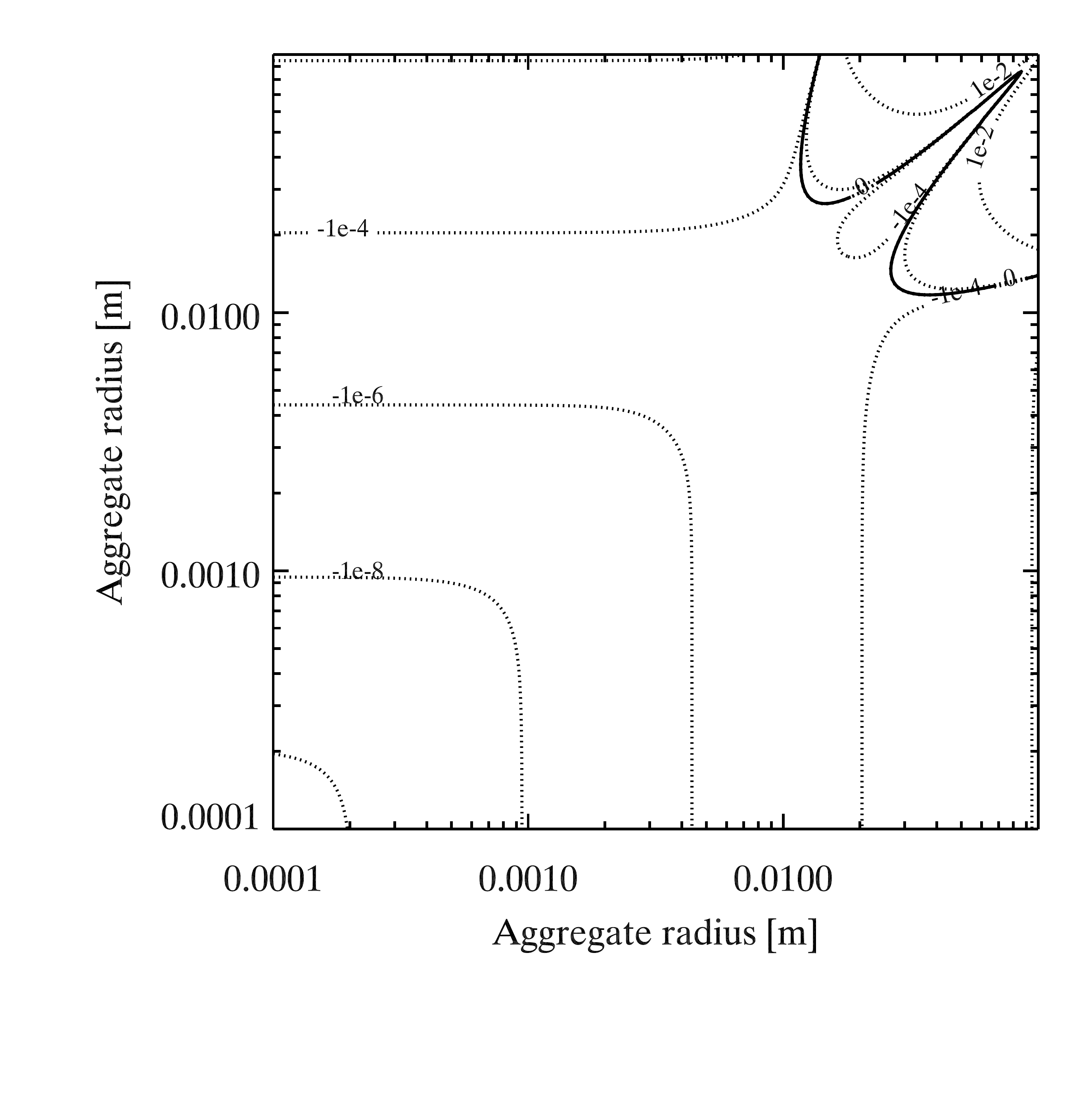}\\
 \caption{Contours of the difference between the predicted collision energy between two clusters ($E_{\textrm{coll}}$) and the fragmentation threshold energy $E_{\textrm{frag}}=3N_{\textrm{c}}E_{\textrm{break}}$ in a Minimum Mass Solar Nebula at 1 AU with $\alpha = 10^{-5}$. The values are given in~J. At the 0 contour line (solid curve) and for positive values, the clusters start losing monomers and begin fragmenting.}
 \label{f:hay_frag}
 \end{center}
\end{figure}

As the relative velocities between clusters composed of $\sim$100 $\mu$m aggregates can be predicted in the protoplanetary disc, their collision energy can be determined as well. If the collision energy $E_{\textrm{coll}}$ is approximated to be only composed of the translational kinetic energy of the colliding aggregate or cluster system, then 
\begin{equation}
E_{\textrm{coll}} = \frac{1}{2}\frac{m_1m_2}{m_1+m_2}v_{\textrm{rel}}^2
\end{equation}
where $m_1$ and $m_2$ are the masses of the colliding clusters and $v_{\textrm{rel}}$ the relative collision velocity. This collision energy can be compared to the threshold energies for the onset of restructuring ($5E_{\textrm{roll}}$) and fragmentation ($3N_{\textrm{c}}E_{\textrm{break}}$) determined in Section \ref{s:recipe}. \\
The minimum energy required for the onset of restructuring was measured to be $E_{\textrm{restr}}$ = $5E_{\textrm{roll}}$ = 8.9$\times10^{-13}$ J. Figure~\ref{f:hay_en} shows the expected collision energy for dust aggregates or clusters in a MMSN model \citep{weidenschilling1977ASS}. The dotted contour indicates the threshold energy for the onset of cluster restructuring measured for M aggregates (see Section \ref{s:recipe} for details). It can be seen that for the chosen disc conditions (at 1~AU with $\alpha = 10^{-5}$), clusters composed of $\sim$100~$\mu$m aggregates would start restructuring at sizes of about 5~mm in radius, marking the onset of cluster compaction (due to restructuring and rolling of the 100~$\mu$m-sized monomers). The dashed contour indicates the threshold energy for the onset of cluster restructuring measured for CM aggregates. For these aggregates, the onset of cluster compaction is at 0.5~mm in radius. This means that the structure of the monomer aggregates constituting the cluster has an influence on the cluster shape. For compacted monomers, clusters restructure into rounder shapes and higher fractal dimensions at smaller monomer sizes (hence earlier during the cluster growth process) while clusters composed of fluffy monomer aggregates keep their fractal structure longer.\\
In addition, for each cluster size, the approximate total number of contacts between the constituting aggregates can be estimated  for compact clusters (i.e. D$_\textrm{f}$=3), as
\begin{equation}
N_{\textrm{c}} = 6\left(\frac{r}{r_0}\right)^3
\end{equation}
 where $r$ is the radius of the cluster, $r_0$ the radius of a monomer aggregate ($\sim$50~$\mu$m) and a mean coordination number of 6, chosen as in Section~\ref{s:recipe}. With the energy required to break one contact between the constituting aggregates of the cluster, which was derived in Section~\ref{s:recipe}, and the sum of the inter-aggregate contacts inside both clusters, the threshold fragmentation energy $E_{\textrm{frag}} = 3N_{\textrm{c}}E_{\textrm{break}}$ can be computed for each pair of cluster sizes, and compared to the collision energy predicted by the protoplanetary disc model. The result of this investigation is presented in Figure~\ref{f:hay_frag}. The values of the difference between the predicted collision energy of two clusters and the fragmentation threshold energy of this collision are given by contours ($E_{\textrm{coll}}-E_{\textrm{frag}}$).  Positive values indicate that the collision energy is high enough to lead to the loss of the first monomers and cluster disruption. In the MMSN model by \cite{weidenschilling1977ASS} (at 1~AU with $\alpha = 10^{-5}$), this happens for clusters of sizes of a few cm. Results for compacted and non-compacted monomer aggregates are very similar (see $E_{\textrm{frag}}$ values in Table~\ref{t:en_macro}).

\section{Conclusion}

We performed a microgravity experiment at the drop tower in Bremen to observe the collision behaviour of $\sim$100~$\mu$m-sized SiO$_2$ aggregates and clusters thereof. We tracked 162 collisions between aggregates and clusters and recorded their collision parameter and outcomes. The velocities at which sticking occurred ranged from 0.18 to 5.0~cm~s$^{-1}$ for aggregates composed of monodisperse dust, with an average value of 2.1$\pm$0.9~cm~s$^{-1}$ for reduced masses ranging from 1.2$\times10^{-6}$ to 1.8$\times10^{-3}$~g with an average value of 2.2$^{+16}_{-2.1}\times10^{-4}$~g. The velocities at which bouncing occurred ranged from 1.9 to 11.9~cm~s$^{-1}$ for the same aggregates with an average of 5.9$\pm$3.2~cm~s$^{-1}$ for reduced masses ranging from 2.1$\times10^{-6}$ to 2.4$\times10^{-4}$ with an average of 7.8$\pm$2.4$\times10^{-5}$~g. The velocities at which fragmentation occurred ranged from 4.9 to 23.8~cm~s$^{-1}$ for the same aggregates with an average of 10.1$\pm$3.2~cm~s$^{-1}$ for reduced masses ranging from 1.2$\times10^{-5}$ to 1.2$\times10^{-3}$ with an average value of 4.2$\pm$2.4$\times10^{-4}$~g. As in previous experiments by \cite{guettler_et_al2010A&A}, \cite{weidling_et_al2012Icarus} and \cite{kothe_et_al2013Icarus}, we observed a transition between sticking and bouncing. However, compared to collisions between individual aggregates, we could see sticking at much higher velocities than expected by these previous models due to the formation of clusters of aggregates during the experiment run, leading to collision events with a mass ratio of up to 10$^4$.\\
The transition between bouncing and fragmentation was also observed to be different from that computed in \cite{kothe_et_al2013Icarus}. We found a transition of the form 
\begin{equation}
\frac{m}{1 \textrm{g}} = 10^{-5.67}\Big[\frac{v}{1 \textrm{cm/s}}\Big]^{1.52}
\end{equation}
where $m$ and $v$ are the reduced mass and relative velocity of the collision, leading to fragmentation at much lower speeds than in previous experiments. As the colliding particles were clusters composed of the initial aggregates, the fragmentation events at lower speeds can be attributed to the fractal nature of the colliding partners.\\
In addition to collisions between free-flying aggregates and clusters, we tracked 101 collisions between clusters and the walls of the experiment cells, observing both restructuring and fragmentation events. We used the collision recipe defined by \cite{dominik_and_tielens1997ApJ} for $\mu$m-sized particles and applied it to our 100~$\mu$m-sized aggregates to measure their critical rolling and breaking energy. We found values of $E_{\textrm{roll}} = 1.8\times10^{-13}$~J and $E_{\textrm{break}}$ = 3.5$\times10^{-13}$~J for non-compacted 100~$\mu$m-sized aggregates. Compacted aggregates showed a similar breaking energy, but a rolling energy 2 orders of magnitude lower. This indicates that cluster restructuring would happen earlier in the particle growth process in the protoplanetary disc if the monomer aggregates were previously compacted \citep[e.g. by bouncing, see ][]{weidling_et_al2009ApJ}. \\
Finally, we applied our results to the particle sizes and velocities expected in current protoplanetary disc models. In a MMSN model \citep{weidenschilling1977ASS}, our results indicate that particles can grow to sizes up to a few millimetres. Cluster restructuring becomes significant for sizes above one millimetre if the monomer aggregates possess a uniform filling factor of $\sim$0.4, while it would start at $\sim$0.3~mm if the monomer aggregates have a compacted rim. The onset of cluster fragmentation takes place for sizes around 30~mm.\\

\markboth{Acknowledgements}{Acknowledgements}
We thank the REXUS/BEXUS project of the Deutsches Zentrum f{\"{u}}r Luft- und Raumfahrt (DLR) for the flight on the REXUS 12 rocket. This work was supported by the ICAPS (Interaction in Cosmic and Atmospheric Particle Systems) project of DLR (grants 50WM0936 and 50WM1236) and a fellowship from the International Max Planck Research School on Physical Processes in the Solar System and Beyond (IMPRS) at  the Max Planck Institute for Solar System Research in G\"ottingen and the University of Braunschweig. We also thank Dipl. Phys. Oliver Lenck from the Frauenhofer Institute for Surface Engineering and Thin Films of Braunschweig for the anti-adhesive glass coating of the particle containers.


\end{document}